\documentclass[11pt,a4paper]{article}
\usepackage[utf8]{inputenc}
\usepackage{jheppub}
\pdfoutput=1

\usepackage{epstopdf, verbatim, ulem, xspace, times, amssymb, mathtools}
\usepackage[table]{xcolor}
\usepackage[nointegrals]{wasysym}
\usepackage{url, textcomp, array, multirow, graphicx, tikz, subfigure}
\usetikzlibrary{shapes,arrows}
\usepackage{hyperref}
\usepackage[titletoc]{appendix}
\newcolumntype{P}[1]{>{\centering\arraybackslash}p{#1}}

\newcommand{\scD}{\mathcal{D}}

\newcommand{\nohatscF}{{\color{white}\hat{\color{black}\mathcal{F}\hspace{-1.2pt}}}}
\newcommand{\scJmunu}{\mathcal{J}^\mu_{\;\,\nu}}
\newcommand{\SMN}{S_M^{\:N}}
\newcommand{\HWC}[1]{{#1}}

\newcommand{\COM}[1]{{}}

\newcommand{\NEW}[1]{{#1}}
\newcommand{\COMR}[1]{{}}

\begin{document}

%%%%%%%%%%%%%%%%%%%%%%%%%%%%%%%%%%%%%%%%%%%%%
%%%%%%%%%%%%%%%%%%%%%%%%%%%%%%%%%%%%%%%%%%%%%

\title{Modification to the Hawking temperature of a dynamical black hole by a \COMR{time-dependent}\NEW{flow-induced} supertranslation}

\author[a,b]{Hsu-Wen~Chiang
}
\author[a,b]{Yu-Hsien~Kung
}
\author[a,b,c,d]{and Pisin~Chen
}

\affiliation[a]{Leung~Center~for~Cosmology~and~Particle~Astrophysics, National~Taiwan~University, \\
Taipei~10617, Taiwan, R.O.C.}
\affiliation[b]{Department of Physics and Center for Theoretical Sciences, National~Taiwan~University, \\
Taipei~10617, Taiwan, R.O.C.}
\affiliation[c]{Graduate~Institute~of~Astrophysics, National~Taiwan~University, \\
Taipei~10617, Taiwan, R.O.C.}
\affiliation[d]{Kavli~Institute~for~Particle~Astrophysics~and~Cosmology, SLAC~National~Accelerator~Laboratory, Stanford~University, \\
Stanford, CA~94305, U.S.A.}

\emailAdd{b98202036@ntu.edu.tw}
\emailAdd{r06222010@g.ntu.edu.tw}
\emailAdd{pisinchen@phys.ntu.edu.tw}

%%%%%%%%%%%%%%%%%%%%%%%%%%%%%%%%%%%%%%%%%%%%%
%%%%%%%%%%%%%%%%%%%%%%%%%%%%%%%%%%%%%%%%%%%%%

\abstract{

One interesting proposal to solve the black hole information loss paradox without modifying either general relativity or quantum field theory, is the soft hair, a diffeomorphism charge that records the anisotropic radiation in the asymptotic region. This proposal, however, has been challenged, given that away from the source the soft hair behaves as a coordinate transformation that forms an Abelian group, thus unable to store any information. To maintain the spirit of the soft hair but circumvent these obstacles, we consider Hawking radiation as a probe sensitive to the entire history of the black hole evaporation, where the soft hairs on the horizon are induced by the absorption of a null anisotropic flow, generalizing the shock wave considered in \cite{Hawking:2016sgy,Chu:2018tzu}. To do so we introduce two different time-dependent extensions of the diffeomorphism associated with the soft hair, where one is the backreaction of the anisotropic null flow, and the other is a coordinate transformation that produces the Unruh effect and a Doppler shift to the Hawking spectrum. Together, they form an exact BMS charge generator on the entire manifold that allows the nonperturbative analysis of the black hole horizon, whose surface gravity, i.e. the Hawking temperature, is found to be modified. The modification depends on an exponential average of the anisotropy of the null flow with a decay rate of 4M, suggesting the emergence of a new 2-D degree of freedom on the horizon, which could be a way out of the information loss paradox.
}

%\arxivnumber{}

\maketitle
\flushbottom

%%%%%%%%%%%%%%%%%%%%%%%%%%%%%%%%%%%%%%%%%%%%%
%%%%%%%%%%%%%%%%%%%%%%%%%%%%%%%%%%%%%%%%%%%%%

\section{Introduction}
\label{sec:intro}

The idea of \HWC{the} black hole thermodynamics, first proposed by Bekenstein \cite{PhysRevD.7.2333}, Bardeen, Carter and Hawking \cite{Bardeen:1973gs}, has been the keystone in the black hole physics. It \COM{formulates the theory of black hole evaporation based on classical thermodynamics}\HWC{paints a picture that black holes evaporate due to their own thermal radiations} \cite{Wald:1993nt}. \COM{Black hole thermodynamics also}\HWC{Such a radiation} plays an essential role in the scrutiny of the quantum theory of gravity due to \COM{the quantum nature of Hawking radiation, which can be described by}\HWC{its quantum origin, i.e.} particle creations near the horizon\HWC{ first realized by Hawking} \cite{Hawking:1974sw}. From the \COM{E}\HWC{e}quivalence principle, \HWC{this }Hawking radiation could be imitated by \COM{considering an observer with a proper acceleration in the Rindler spacetime}\HWC{a properly accelerating observer described by the Rindler coordinate} \cite{PhysRevD.7.2850,Davies_1975,PhysRevD.14.870}\COM{. The latter refers to another physical process, the Unruh effect, that an accelerated observer would experience additional black body radiation \COM{as compared with another}\HWC{when compared to an} inertial observer in the same flat spacetime \cite{PhysRevD.14.870}. Calculating the particle number counted by the accelerated observer by using Bogoliubov transformation, one can obtain a thermal spectrum with an absolute temperature which is called Unruh temperature. Combining with the derivation of Hawking temperature, this gives us a subtle relation that}\HWC{, subtly hinting that the associated temperature, i.e.} the Hawking\COM{-Unruh} temperature could be \COM{represented by}\HWC{related to} the surface gravity \COM{$\kappa$ }\HWC{on the horizon.}%arising from an in-affine parameterized curve. \COM{The minimal conditions for the existence of Hawking radiation have recently been found}\HWC{Recently the minimal conditions for the existence of Hawking radiation and especially the  relation between the Hawking temperature and the null geodesic congruences are detailed} in \cite{PhysRevD.83.041501}.

%The zeroth law of black hole thermodynamics states that the surface gravity of a stationary black hole is a constant which can be imitated by a thermal equilibrium system sharing a constant temperature. More precisely, an observer in an asymptotically flat region observes a constant temperature $\kappa/2\pi$. The surface gravity could be defined by the Killing horizon for a stationary black hole. The Killing vector fields $\xi$ normal to the Killing horizon satisfy $\xi^a \nabla_a \xi^b = \kappa \xi^b$ \cite{Nielsen_2008}. Independent of the Einstein field equation, the derivation of Hawking radiation is valid for arbitrary spacetimes and the zeroth law should hold generally at least for stationary black holes.
Another central property of the black hole is that a stationary black hole has no hair, i.e. no parameter other than the mass, the angular momentum and the charges \cite{1967PhRv..164.1776I}. For non-stationary black holes it has been shown \cite{Dafermos:2013bua,PhysRevLett.123.111102} that the hairs are quickly lost even if successfully implanted. This so-named no-hair theorem (stemming from the Einstein field equation), when combined with the Hawking radiation (depending only on the quantum field theory in the curved spacetime), severely challenges our understanding of both theories. Despite the classical black hole carrying no entropy, by the thermality of the Hawking radiation some entropy is bound to be generated, and nowhere can that entropy be mitigated unless the horizon is bypassed. This is the information loss paradox \cite{PhysRevD.14.2460,Mathur_2009}, a direct violation of the unitarity, i.e. the foundation of the quantum mechanics used to derive the Hawking radiation in the first place. Thus an inevitable conclusion is reached that one of the three tenets: the no-hair theorem, the locality, the quantum mechanics, has to be forfeited \cite{Almheiri:2012rt}.

%However, there still exists a contradiction between black hole thermodynamics and the no-hair theorem in general relativity. Therefore, the black hole information paradox \cite{Mathur_2009} arises consequently. The in-falling pure states that are maximally entangled become mixed states in the form of Hawking radiation in the evaporation process of a black hole. Due to the no-hair theorem, it is well-known that the information is limited outside the horizon. The mixed thermal states cannot encode all the information of the pure states and the unitarity in quantum mechanics is thus violated.
Several candidates of\HWC{ the} resolution to \COM{this}\HWC{the} paradox have been proposed\COM{ in the past} \cite{Almheiri:2012rt,Chen:2014jwq,Unruh:2017uaw,Maldacena:2013xja}. %researches.
One particular proposal, the black hole soft hair by Strominger has attracted attentions recently \cite{Hawking:2016msc,Strominger:2013jfa,Strominger:2017zoo}, where the soft hair, i.e. the conserved charge of the BMS symmetry \cite{Bondi:1962px,Sachs:1962wk,PhysRev.128.2851}, a residue of the diffeomorphism associated with the asymptotic Killing vector in asymptotically flat spacetimes, would serve as the entropy storage.
%
%Recently, the idea of the soft hair of black holes, a possible resolution proposed by Hawking, Perry, and Strominger, has been widely studied \cite{Hawking:2016msc,Strominger:2013jfa,Strominger:2017zoo}. Their research is based on the BMS symmetry in asymptotically flat spacetimes \cite{Bondi:1962px,Sachs:1962wk,PhysRev.128.2851}, which is found as an spacetime diffeomorphism. A degree of freedom, the supertranslation, \COM{appears as the}\HWC{an} angle-dependent \COM{shift of the time coordinate}\HWC{time translation} in the asymptotically flat region. The BMS symmetry is an infinite-dimensional symmetry which transforms an asymptotically flat solution to another degenerate solution. When a spacetime contains an isolated black hole,
\HWC{It would require} the asymptotic region near the event horizon \COM{would}\HWC{to} exhibit \COM{another}\HWC{a} copy of BMS symmetry\HWC{, and a channel through which the entropy on the horizon could be released, such that after the complete evaporation of the black hole the unitarity could be preserved}.% After the black hole is completely evaporated, the data on different Cauchy surfaces would be preserved, and the information is not lost.
\HWC{ Such a symmetry was discovered in \cite{Hawking:2016msc,Hawking:2016sgy}, where the soft hair is successfully implanted at the linear order on the horizon of a Schwarzschild black hole by an incoming anisotropic shock wave focused on the central singularity, leaving only the covert channel to be found.}

%With the antipodal matching condition studied by Christodoulou and klainerman \cite{Christodoulou:1993uv}, the BMS charge tends to be conserved in the region between past null infinity and future null infinity. If we know the material forming a black hole that came from past null infinity, then we could expect that at the future null infinity, a certain amount of information might be encoded in Hawking radiation. Then the infinite amounts of BMS charges could be generated as a new kind of black hole hair \cite{Hawking:2016msc,Hawking:2016sgy} that stores the information in the black hole evaporation process. Besides, from a classical point of view, the soft hair is still consistent with no-hair theorem since it preserves the spacetime diffeomorphism after BMS transformations. The physical observable is expected to be detected through the gravitational memory effect. 

%On the other hand, the issue about whether the soft modes could constrain hard modes in the scattering process has been addressed by using “dressing” factorization \cite{Bousso:2017dny,PhysRevD.96.086016} . The factorization leads to the decoupling between soft and hard modes. The motivation of our work is neither considering the quantum nature of hard mode scattering nor the study of unitarity. Instead, we will focus on the classical feature of black hole thermodynamics and figure out whether the supertranslation could have a nontrivial effect on Hawking temperature.

While the soft hair proposal may sound pretty convincing, there are several obstacles ahead, mostly related to how soft hairs interact and release the stored entropy. One issue is the inability to measure the soft hair within the BMS symmetry group itself as it is abelian, indicating that we may not be able to measure the soft hair of a black hole directly from afar. People have since been trying to enhance the symmetry \cite{Barnich:2011mi,Pasterski:2015tva}. Meanwhile, it became apparent that it is exceedingly hard to discern the soft hair from zero-frequency gravitational waves (soft gravitons) \cite{Bousso:2017dny,PhysRevD.96.086016}. At the infrared limit, the off-shell graviton generating the BMS symmetry is indistinguishable from the on-shell graviton, and should not be considered as a standalone observable. By the factorization procedure one may decouple the soft particles from the hard (non-zero-frequency) particles constituting the BMS charges, and thus entirely negate the purpose of the soft hair.

To overcome the difficulties one should consider a nonlocal measurement that depends on the near-black-hole geometry explicitly. One candidate would be the Hawking radiation, the origin of all the hassle. Unfortunately in \cite{Hawking:2016sgy}, the Hawking temperature was shown to be insensitive to the shock-wave-induced soft hair on a Schwarzschild black hole, at least at the linear order when away from the shock wave. Furthermore in \cite{Javadinazhed:2018mle,Comp_re_2019} by utilizing the dressed state, the decoupled hard particle in the factorization procedure, the modification to the Hawking radiation spectrum by the soft hair was derived and found to be merely a phase shift. These negative results are not surprising given the lack of dynamics to distinguish soft hairs from soft gravitons. To introduce more dynamics, in \cite{Chu:2018tzu} a Vaidya black hole was considered and a small perturbation to the surface gravity and the Hawking temperature was found. However as we will explain in section \ref{sec:HawkingTemp}, it is just an incarnation of the diffeomorphism that BMS symmetry belongs to.

%In \cite{Hawking:2016sgy}, the soft hair was assumed to be generated by a shock wave in Schwarzschild spacetime. According to this setup, the Hawking temperature was shown to be insensitive to the \HWC{shock-wave-induced} soft hair. A \COM{similar}\HWC{more rigorous} result was \COM{also found}\HWC{derived} in \cite{Javadinazhed:2018mle,Comp_re_2019} by the dressed variable formalism. The Hawking spectrum is merely modified by a phase shift and the Hawking temperature remains intact. Based on these studies, Koyama and Chu \cite{Chu:2018tzu} have studied the soft hair in dynamical spacetimes with an implanted shock wave. By applying the simplest form of dynamical spacetime, the Vaidya spacetime, they found a modification of surface gravity in $O(f)$ and it will be discussed in the following section. 

%Another major reason to consider such a setup would be the parallel between the incoming matter forming the black hole and the outgoing Hawking radiation evaporated from the black hole. If the imposed energy flow could generate any geometric information, we expect the Hawking radiation will also generate the same effect during the black hole evaporation process, thus possibily shedding new lights on the information loss paradox.

Still, this is expected given that the necessity of dynamics refers to the soft hair, rather than the background in the case of the Vaidya spacetime. Therefore in this work, we will generalize the generators of the BMS symmetry to incorporate the dynamics necessary to distinguish between the ``not-so-soft'' hair induced by an incoming continuous anisotropic null flow, the diffeomorphism for the dressing procedure, and the ``not-so-soft'' gravitons.
%In this work, we will show that by only considering a dynamical background is not sufficient to give non-trivial modification on surface gravity. The surface gravity would remain the same form after applying a $O(f)$ coordinate transform on the whole metric. That is, under all BMS transformations, the Hawking temperature will still be $\kappa / 2\pi$. Our setup is to generalize the angle function $f\left(\Theta\right)$ to be a time-dependent function $f\left(\tau,\Theta\right)$ that represents the backreaction on the background spacetime of a continuous energy flow. We will apply this assumption to both the transformation of the dynamical background and the dressing factorization. With these new kinds of transformations with \COMR{time-dependent}\NEW{flow-induced} supertranslation, there are two cases that should be distinguished, according to whether the spacetime remains diffeomorphic.
As \COM{we will }exhibit\HWC{ed} later, the \HWC{``not-so-soft'' hairs}\COM{case that breaks the diffeomorphism would} cause a \HWC{temporal but }nontrivial effect on\HWC{ the} surface gravity. %As a consequence, by introducing the \COMR{time-dependent}\NEW{flow-induced} supertranslation, the modification of Hawking temperature seems to resolve
\HWC{Notice that these setups contain only the incoming flow, thus excluding the back-reaction of the Hawking radiation necessary for a self-consistent picture. Nevertheless given the tie between the Hawking radiation and its negative-energy partner falling into the black hole, the modification could still shed}%elevate} the contradiction between black hole thermodynamics and the no-hair theorem, shedding 
some light on \COM{ameliorating }the black hole information problem.

The organization of the paper is as follows. In section \ref{sec:transforms}, we briefly review the Bondi-Sachs formalism that \COM{contains the}\HWC{is} essential \COM{materials }for our analysis\HWC{, and introduce two different transformations related to the dynamical soft hair}. %We introduce four kinds of Vaidya spacetimes with supertranslation. These different setups determine the modification of Hawking temperature.
In section \ref{sec:dressing}, we review the dressing \COM{factorization }in \cite{Bousso:2017dny,Javadinazhed:2018mle} and generalize it to a \COM{time-dependent}\HWC{dynamical} scenario. In section \ref{sec:main}, we \HWC{relate the Hawking temperature to the surface gravity in a generic spacetime and demonstrate that the surface gravity could be non-covariantly modified by the null flow. }%calculate the surface gravity in different setups of dynamical spacetimes. The modification \COM{of}\HWC{to the} Hawking temperature could be explicitly demonstrated. 
In section \ref{sec:discussion}, we discuss the physical implication\HWC{ and possible extension} of our works.% and the relation to the information loss paradox.

%%%%%%%%%%%%%%%%%%%%%%%%%%%%%%%%%%%%%%%%%%%%%
%%%%%%%%%%%%%%%%%%%%%%%%%%%%%%%%%%%%%%%%%%%%%

\section{Soft hairs on dynamical black holes}
\label{sec:transforms}

%The recent studies \cite{Strominger:2013jfa,Hawking:2016sgy} indicates that soft hairs might preserve the information of the Hawking radiation. However given the reasons in the previous section, we need a dynamical soft hair to elevate the soft factorization \cite{Bousso:2017dny,PhysRevD.96.086016,Javadinazhed:2018mle,Comp_re_2019}.
In this section, we will first introduce the BMS metric in the advanced Bondi coordinate, where the vanilla soft hair is found as the conserved charge of the residue diffeomorphism on the past null infinity $\mathcal{I^-}$, i.e. the BMS symmetry. For simplicity, we will not venture into the issue of the other BMS symmetry on the future null infinity $\mathcal{I^+}$. Instead, we will focus on the matter falling into the future horizon $\mathcal{H}$ that may model the gravitational collapse and mimic the Hawking radiation partners. We will then discuss the shock-wave-induced soft hair \cite{Hawking:2016sgy,Chu:2018tzu}, and generalize it to be time-dependent. In the process, we realize the existence of another type of the covariant transformation, previously mistaken as merely the transformation within the BMS group. These two together form the foundation for further discussions in this work.
%Our purpose is to consider the modification to the Hawking radiation in different setups of dynamical spacetimes. The simplest model \COM{of the dynamical spacetime is}\HWC{would be} the Vaidya spacetime \COM{which allows the time evolution of the}\HWC{with a time-dependent} Bondi mass. In the Bondi coordinate, there are two sets of null hypersurfaces corresponding to past null infinity $\mathcal{I^-}$ and future null infinity $\mathcal{I^+}$. The Hawking radiation is emitted at the future horizon $\mathcal{H}$ of the black hole and propagates to $\mathcal{I^+}$. The recent studies \cite{Strominger:2013jfa,Hawking:2016sgy} indicates that the \COM{memory effect}\HWC{soft hairs} in the asymptotic\COM{ally flat} region might preserve the information of the Hawking radiation by supertranslation soft hairs. Our study is based on the region described by the advanced Bondi coordinates where the matter falls into the black hole.

\subsection{Asymptotic symmetry on an asymptotically Minkowski spacetime}
\label{sec:BMS}

In the seminal work by Bondi, Van der Burg, Metzner and Sachs (BMS) \cite{Bondi:1962px,Sachs:1962wk,PhysRev.128.2851}, the authors found that an asymptotically Minkowskian region can be represented by a family of metrics with appropriate fall-off conditions. In this region, one can impose different fall-off conditions depending on the physical situations under consideration. The constraints should be loose enough to contain the non-trivial solutions such as gravitational waves but strict enough to rule out unphysical ones.

Considering an asymptotically Minkowski region $\mathcal{I}^-$ along the past null direction on a $(\text{3+1})-$D manifold, one can introduce the advanced coordinates $\left(v, r, \Theta^A\right)$ in the Bondi gauge, where $v$ is the advanced time, $r$ the areal radius, and $\Theta^A$ the coordinates of a unit 2-sphere $S^2$. Following the notation in \cite{Hawking:2016sgy}, the gauge condition and the metric up to next-next-leading order in $1/r$ are
\begin{align}
g_{rA} = g_{rr} = 0  \;\,,\quad
\det \big( g_{AB} \big) = \det \big( r^2 \gamma_{AB} \big)  \:,  \label{eq:BMSgauge}
\end{align}
\begin{align}\SwapAboveDisplaySkip
ds^2 &= - dv^2 + 2dv dr + r^2 \gamma_{AB} d\Theta^A d\Theta^B  \nonumber \\
     &+ \frac{2m}{r} dv^2 + r C_{AB} d\Theta^A d\Theta^B - \scD^B C_{AB} dv d\Theta^A - \frac{1}{16r^2} C_{AB} C^{AB} dv dr  \nonumber \\
     &- \frac{4}{3r} \left( N_A - v \partial_A m - \frac{3}{32} \partial_A \left( C_{BD} C^{BD} \right) \right) dv d\Theta^A + \frac{1}{4} \gamma_{AB} C_{DE} C^{DE} d\Theta^A d\Theta^B  \nonumber \\
     &+ \ldots  \,,  \label{eq:gBMS}
\end{align}
where $\gamma_{AB}$ is the metric of $S^2$, $m$ the Bondi mass aspect, $N_A$ the angular momentum aspect, and $C_{AB}$ a traceless tensor (shear). The latin indices $A$, $B$, $C$ are lowered and raised by $\gamma$.
The asymptotically flat condition and the constraint equations for the dynamical variables $\left(m, N_A, C_{AB}\right)$ are
\begin{align}
\partial_r m =&\; \partial_r C_{AB} = \partial_r N_A =0  \;\,,\\
%N_{AB}  \equiv&\; \partial_v C_{AB}  \;\,,\\
\partial_v m =&\; \frac{1}{4} \scD^A \scD^B N_{AB} + \frac{1}{8} N_{AB} N^{AB} + 4\pi r^2 T_{vv} \Big{|}_{\mathcal{I}^-}  \label{eq:Tvv}  \,,\\
\partial_v N_A =&\; \frac{1}{4} \epsilon_{AB}\epsilon^{DE} \Big( \scD^B \scD_E \scD^F C_{FD} - \scD_E \left( C^{BF} N_{FD} \right) \Big)  \nonumber \\
+&\; \frac{1}{2} C_{AB} \scD_D N^{BD} + v \partial_A \partial_v m - 8\pi r^2 T_{vA} \Big{|}_{\mathcal{I}^-}  \,, \label{eq:TvA}
\end{align}
where $T_{\mu\nu}$ is the matter stress tensor, $N_{AB} \equiv \partial_v C_{AB}$ the Bondi news and $\scD$ the covariant derivative projected onto $S^2$. Despite the choice of the Bondi gauge, these dynamical variables are not unique and are related by a local time translation on the 2-sphere, i.e. supertranslation,
\begin{align}
\delta v = f  \:\,,\quad
\delta r = - \frac{1}{2} \scD^2 f + O\left(r^{-1} \right)  \:,\quad
\delta \Theta^A = \frac{1}{r} \scD^A f + O\left(r^{-2} \right)  \:,
\end{align}
where $f$ is an arbitrary function on $S^2$. Together with the Lorentz transformation they form the BMS transformation. The associated generating vector field is
\begin{align}
\zeta_f = f \partial_v - \frac{1}{2} \scD^2 f \partial_r + \frac{1}{r} \scD^A f \partial_A  \label{eq:generator}  \:\,.
\end{align}

%Notice that $\zeta_f$ generates both the residual diffeomorphism after gauge fixing according to eq.~\eqref{eq:BMSgauge}, and the only direction

Following the same procedure there would be another copy of BMS transformation on another asymptotically Minkowski region $\mathcal{I}^+$ along the future null direction. However as first shown by Christodoulou and Klainerman \cite{Christodoulou:1993uv} and later reinterpreted by Strominger \cite{Strominger:2013jfa}, two transformations should be related by the antipodal matching condition at the spatial infinity $i^0$ to preserve the strong asymptotically flat condition that guarantees their co-existence. This relation halves the amount of symmetries in the gravitational scattering process.

\subsection{Supertranslated Vaidya metric}
\label{sec:oldtransform}
First shown in \cite{Hawking:2016msc} that the $U(1)$ version of the BMS transformation can be applied to a charged static black hole and naively extended to its horizon without any issue, Hawking, Perry and Strominger further demonstrated \cite{Hawking:2016sgy} that the supertranslation of a Schwarzschild black hole can be actively generated by an incoming light-like shock wave. Unfortunately in \cite{Javadinazhed:2018mle} the authors proved that the effect of the supertranslation on the Hawking radiation or any other physical observable is insensible in the case of the Schwarzschild spacetime. Therefore we have to consider more general setups, e.g. the Vaidya spacetime consisting of an isotropically accreting black hole, which will serve as the basis of all the derivatives in this work.

In the advanced Bondi coordinates the Vaidya spacetime can be written as
\begin{align}
ds^2 = g^{\!\text{Vaidya}}_{\mu\nu} dx^\mu dx^\nu = -Vdv^2 + 2dvdr + r^2 \gamma_{AB} d\Theta^A d\Theta^B  \,,\quad
V \equiv 1 - \frac{2M}{r}  \:\,,
\end{align}
where the mass aspect $M$ only depends on $v$. The associate energy momentum tensor is
\begin{align}
T_{vv}^\text{Vaidya} = \frac{M'}{4 \pi r^2} \equiv \frac{\partial_v M}{4 \pi r^2}  \:\,.
\end{align}
From now on we denote $\partial_v$ by the prime, and drop the superscript ``Vaidya'' as the Vaidya spacetime will be the basis of all transformations. After the supertranslation the metric becomes
\begin{align}
ds^2 = &- \left( V - \frac{2 M' f}{r} - \frac{M \scD^2 f}{r^2} \right) dv^2 + 2 dvdr - \scD_A \left( \scD^2 f + 2V f \right) dv d\Theta^A  \nonumber\\
       &+ \left( r^2 \gamma_{AB} + 2r \scD_A \scD_B f - r \gamma_{AB} \scD^2 f \right) d\Theta^A d\Theta^B  \label{gBMSV}  \,.
\end{align}
The energy momentum tensor is transformed accordingly as
\begin{align}
\label{eq:BMStransT}
T_{\mu\nu} \rightarrow T_{\mu\nu} + \mathcal{L}_f T_{\mu\nu}  \;\,,
\end{align}
where $\mathcal{L}_F$ is the Lie derivative with respect to the vector field defined in eq.~\eqref{eq:generator}, with the function $f$ being replaced by another function $F$. The transformation-induced anisotropies are
\begin{align}
\mathcal{L}_f T_{vv} = \frac{M'' f}{4\pi r^2} + \frac{M' \scD^2 f}{4\pi r^3}  \:\,,\quad
\mathcal{L}_f T_{vA} = \frac{M' \scD_A f}{4\pi r^2}  \label{eq:tST}  \:\,.
\end{align}

However notice that a supertranslation is still within the reach of \cite{Javadinazhed:2018mle}. We need a more dynamical system. In \cite{Chu:2018tzu} the authors consider a setup with a light-like shock wave falling into the black hole, similar to that of \cite{Hawking:2016sgy}, except replacing the background Schwarzschild spacetime with the Vaidya spacetime. The resulting shock-wave-induced supertranslation (SST) can be written as
\begin{align}
g_{\mu\nu} \rightarrow g_{\mu\nu} + \theta\left(v-v_0\right) \mathcal{L}_f g_{\mu\nu} %\Big{|}_{f \rightarrow \theta\left(v-v_0\right) f}
\label{eq:gSST}  \;\,,
\end{align}
where $\theta\left( v - v_0 \right)$ is the Heaviside theta function. This transformed metric describes two Vaidya spacetimes, one vanilla and another supertranslated by $f$, cut and glued together at $v=v_0$.

To find out the content of the shock wave, we extract terms with the Dirac delta function $\delta\left(v-v_0 \right)$ from the energy momentum tensor as the following:
\begin{align}
T_{\mu\nu} &\rightarrow  %T^{SST} &=
T_{\mu\nu} \Big[ g_{\mu\nu} + \theta\left(v-v_0\right) \mathcal{L}_f g_{\mu\nu}   %\Big{|}_{f \rightarrow \theta\left(v-v_0\right) f}
\Big]  \label{eq:TSST}  \;,\\
T_{vv}^{\delta} &= \left( -\frac{1}{4} \scD^2 \left( \scD^2 +2 \right) f + M' f + \frac{3M}{2r} \scD^2 f
\right) \frac{\delta\left(v-v_0 \right)}{4\pi r^2}  \label{eq:SSTvv}  \:\,,\\
T_{vA}^{\delta} &= \frac{3M}{2} \scD_A f \frac{\delta\left(v-v_0 \right)}{4\pi r^2}  \:\,,  \label{eq:SSTvA}
\end{align}
where $T_{\mu\nu} \big [ q_{\mu\nu} \big ]$ is the energy momentum tensor derived from a metric $q_{\mu\nu}$ and $T_{\mu\nu}^{\delta}$ is the perceived shock wave. From the response theory point of view, the ``shock wave'' from the junction condition  %these additional terms in the energy momentum tensor
eqs.~\eqref{eq:SSTvv} and \eqref{eq:SSTvA} activates the supertranslation when passing through at $v =v_0$.

We may compare these  %forms
with the  %ansatz given in \cite{Chu:2018tzu}. The 
anisotropic part of the energy momentum tensor %of the implanted shock wave defined
given in \cite{Chu:2018tzu}\footnote{We only consider $\mu =0$ case in \cite{Chu:2018tzu}.}
\begin{align}
T^\text{SST}_{vv} &= \frac{1}{4\pi r^2} \Big( \hat{\mu} + T \delta \left(v-v_0 \right) \Big) + \frac{1}{4\pi r^3} \Big( T^{\left(1 \right)} \delta \left(v-v_0 \right) + t^{\left(1 \right)} \theta \left(v-v_0 \right) \Big)  \nonumber  \;,\\
T^\text{SST}_{vA} &= \frac{1}{4\pi r^2}  \Big( T_A \delta \left(v-v_0 \right) + t_A \theta \left(v-v_0 \right) \Big)  \:,
\end{align}%\vspace{-.038\textwidth}
where
\begin{align}\SwapAboveDisplaySkip
\qquad\qquad
T &= - \frac{1}{4} \scD^2 \left( \scD^2 +2 \right) f  \:\,,&
T^{\left(1 \right)} &= \frac{3M}{2} D^2 f  \:\,,&\;
t^{\left(1 \right)} &= M' \scD^2 f  \nonumber  \:\,,\qquad\qquad\quad\,\\
\hat{\mu} &= \partial_v \Big( \theta \left(v-v_0 \right) M' f \Big)  \:,&
T_A &= \frac{3M}{2}  \scD_A f  \:\,,&\;
t_A &= M' \scD_A f  \:\,.
\end{align}
Apparently the energy momentum tensor found in \cite{Chu:2018tzu} includes Heaviside theta terms induced by the \COMR{BMS transformation}\NEW{supertranslation} as shown in eqs.~\eqref{eq:tST}. After subtracting those terms, we are left with terms equivalent to eqs.~\eqref{eq:SSTvv} and \eqref{eq:SSTvA}
\begin{align}
T_{vv}^{\delta} = \frac{1}{4\pi r^2} \Big( \bar{\mu} + T \delta \left(v-v_0 \right) \Big) + \frac{1}{4\pi r^3} T^{\left(1 \right)} \delta \left(v-v_0 \right)  \:,\quad
T_{vA}^{\delta} = \frac{1}{4\pi r^2} T_A \delta \left(v-v_0 \right)  \:,
\end{align}%\vspace{-.038\textwidth}
where
\begin{align}\SwapAboveDisplaySkip
\bar{\mu} = \delta \left(v-v_0 \right) M' f  \:\,.  \label{eq:mubar}
\end{align}
The form of the energy momentum tensor above is the same as that the Schwarzschild black hole, which shouldn't be too surprising given eqs.~\eqref{eq:Tvv} and \eqref{eq:TvA}'s validity near the event horizon of the Schwarzschild black hole as shown in \cite{Hawking:2016sgy}.

However, even though the subtracted terms appear to be related to the \COMR{BMS transformation}\NEW{supertranslation}, they do not stem from any proper coordinate transformation. An actual coordinate transformation\NEW{, without/with a supertranslation before/after the shock wave has passed, should be} generated by \NEW{a piecewise vector field }$\theta \left(v-v_0 \right) \zeta_f$ would spawn an additional term $- \delta \left(v-v_0 \right) M' f / \left( 2\pi r^2 \right)$ for $T_{vv}$. Eqs.~\eqref{eq:SSTvv} and \eqref{eq:mubar} should be corrected accordingly and only then the resulting energy momentum tensor $T_{vv}^S$ can be regarded as the the shock wave.

\subsection{Two different extensions to the time-independent supertranslation}
\label{sec:tdtransform}
In this subsection we will extend respectively the shock-wave-induced supertranslation and the associated time-dependent coordinate transformation introduced in the last paragraph to describe the anisotropic continuous null flow.

The first kind of the transformation is called ``time-dependent \COMR{BMS}\NEW{supertranslation''} (\COMR{tBMS}\NEW{tST})\COMR{ transformation''} \COMR{where the parameter $f$ of}\NEW{which is a generalization to} the generating vector field \NEW{defined in} eq.~\eqref{eq:generator} \NEW{by adding time dependency to the parameter $f$}\COMR{ is generalized to be time-dependent}. \COMR{The tBMS transformation}\NEW{TST}  \COMR{is}\NEW{as} a coordinate transformation \NEW{on $\mathcal{I}$}\COMR{that} preserves all physical observable up to a non-trivial Bogoliubov transformation (c.f. section \COMR{\ref{sec:bogo}}\NEW{\ref{sec:dressing_tST}})\NEW{, and may be regarded as a generalization to the supertranslation in the same way as the isothermal coordinate of flat spacetime is to the Minkowski metric (c.f. \cite{1981PhRvD..24.2100S}), with the usual constraint of invertability on $\mathcal{I}$, i.e. the monogamy of the time and spherical coordinates $f'(v)<1$ and $\left(\scD^2 +2\right)f'(v) <1$. Furthermore, in the same way as the piecewise generating vector field introduced in the previous subsection is essential for the understanding of SST, tST is also necessary for the second kind of the transformation about to be introduced below. For more details about the applicability of this new coordinate transformation please refer to section 5}.

The second kind is called ``\COMR{time-dependent}\NEW{flow-induced} supertranslation'' (\COMR{tST}\NEW{FST}) which \COMR{similar to eq.~\eqref{eq:gSST} is not a coordinate transform, but a supertranslation actively}\NEW{generalizes eq.~\eqref{eq:gSST} by substituting the piecewise supertranslation generated by a incoming shock wave, with a time-dependent one} generated by \COMR{the}\NEW{an} incoming null flow. It transforms the metric in exactly the same way as the time-independent supertranslation, except with a time-dependent parameter $f (v,\Theta^A)$. Since after its application the spacetime is not diffeomorphic to the original, this transformation allows us to construct various black hole systems with rich dynamics and modified surface gravities, which will be discussed in section \ref{sec:main}. \NEW{Furthermore, as shown in \cite{Christodoulou:1993uv} FST is the only form for the sub-leading part of the initial condition at $\mathcal{I}^-$ that leads to a stable asymptote, and thus is the most natural extension one would consider after SST in \cite{Hawking:2016sgy}.}

We will demonstrate that these two are highly correlated, and together they form a linear response function between the anisotropic incoming null flow and the transformation parameter $f$.

%%%%2.3.1

%\subsubsection{time-dependent \COMR{BMS transformation}\NEW{supertranslation}}
According to the definition above, \COMR{the tBMS transformation}\NEW{tST} is obtained by substituting the time-independent function $f(\Theta )$ in the generating vector field $\zeta_f$ by a time-dependent one $f(v,\Theta )$ (denoted as $f(v)$ for simplicity). The resulting coordinate transformation can be written as
\begin{align}
g_{\mu\nu} \rightarrow g_{\mu\nu} + \mathcal{L}_{f\left(v\right)} g_{\mu\nu}  \;\,.
\end{align}
The \COMR{tBMS}\NEW{tST} transformed \NEW{(tSTed) }metric of the Vaidya spacetime becomes
\begin{align}
ds^2 &= - \left( V - \frac{2M' f}{r} - \frac{M \scD^2f}{r^2} + \left(\scD^2+2V\right) f' \right)dv^2 + 2\left(1+f'\right)dvdr  \label{eq:gtST}  \\
&- \scD_A \left( \scD^2 f + 2V f + 2rf'\right)dvd\Theta^A + \left(r^2 \gamma_{AB} +2r \scD_A \scD_B f - r \gamma_{AB} \scD^2 f \right) d\Theta^A d\Theta^B  \,.  \nonumber
\end{align}
Compared with eq.~\eqref{gBMSV}, the additional terms are clearly due to the time dependence of \COMR{the tBMS transformation}\NEW{tST}, as they are all proportional to $f'$ and vanish as we recovers the original \COMR{BMS transformation}\NEW{time-independent supertranslation}. Notice that \COMR{tBMS transformation}\NEW{tST} is a coordinate transformation, and its effect on the energy momentum tensor can be obtained simply by the covariant transformation
\begin{align}
T_{\mu\nu} \rightarrow  T_{\mu\nu} + \mathcal{L}_{f\left(v\right)} T_{\mu\nu} \;\,.
\end{align}
The induced anisotropies of the energy momentum tensor are
\begin{align}
T_{vv}^\text{\COMR{tBMS}\NEW{tST}} = \frac{M' f'}{2\pi r^2} + \frac{M'' f}{4\pi r^2} + \frac{M' \scD^2 f}{4\pi r^3}  \:\,,\quad
T_{vA}^\text{\COMR{tBMS}\NEW{tST}} = \frac{M' \scD_A f}{2\pi r^2}  \:\,.
\end{align}
%%%%2.3.2

%\subsubsection{\COMR{time-dependent}\NEW{flow-induced} supertranslation}

Parallelly as the generalization to eq.~\eqref{eq:gSST}, the \COMR{time-dependent}\NEW{flow-induced} supertranslation (\COMR{tST}\NEW{FST}) describes the process of an anisotropic continuous flow falling into the black hole while inducing varying amounts of supertranslation. It is defined %in a similar way as eq.~\eqref{gSST}
 as a non-covariant transformation on the metric
\begin{align}
g_{\mu\nu} &\rightarrow g_{\mu\nu} + \mathcal{L}_f g_{\mu\nu} \Big{|}_{f \rightarrow f\left(v\right)}  \:.  \label{eq:FST}
\end{align}
The corresponding energy momentum tensor becomes
\begin{align}
T_{\mu\nu} \rightarrow T_{\mu\nu} + T_{\mu\nu}^\text{\COMR{tST}\NEW{FST}} = T_{\mu\nu} \left[ g_{\mu\nu} + \mathcal{L}_{\zeta_f} g_{\mu\nu} \Big{|}_{f \rightarrow f\left(v \right)} \right]  \,,
\end{align}
where $f \rightarrow f(v)$ denotes the replacement of the time-independent $f$ by a time-dependent $f(v)$ after applying the Lie derivative, and $T^\text{\COMR{tST}\NEW{FST}}$ is the anisotropic part of the energy momentum tensor.

The resulting metric appears exactly the same as that of eq.~\eqref{gBMSV} except with a time-dependent $f$, and is quite different from that of eq.~\eqref{eq:gtST}. The difference leads to an additional anisotropic null flow, which can be written as an time integration over the shock wave at $v_0$
\begin{align}
T_{\mu\nu}^\text{\COMR{tST}\NEW{FST}} &- T_{\mu\nu}^\text{\COMR{tBMS}\NEW{tST}} =
\int_{v_0=-\infty}^{v_0=v} T_{\mu\nu}^S \left( v_0\right)  \nonumber  \:,\\
T_{\mu\nu}^S \left( v_0\right) \equiv \big( T_{\mu\nu}^\text{\COMR{tST}\NEW{FST}} &- T_{\mu\nu}^\text{\COMR{tBMS}\NEW{tST}} \big) \Big{|}_{f \left(v \right) \rightarrow f' \left(v_0 \right) \theta \left(v-v_0\right)}  \;.  \label{eq:responseTf}
\end{align} 
where $T_{\mu\nu}^S$ is the shock wave energy content introduced at the end of section \ref{sec:oldtransform}.
\NEW{While one can treat $T^S$ as the part of the anisotropy actually responsible for the deformation of the manifold, and $T^\text{tST}$ as a compensation to the backreaction on the background flow $M'$ by the metric response in the past, given that the backreaction can be cast into a coordinate transformation it is more natural to consider the above equation as the evidence of a special gauge reachable from the Bondi gauge by a tST transformation at the linear order, where the response relation between the energy momentum tensor anisotropy and the metric deformation becomes linear.}\COMR{Therefore the two transformations introduced in this section together form a linear response relation between the energy momentum tensor of the in-falling matter and the transformation function $f(v)$.} The combined transformation clearly has nice and clean properties that we will discuss in section \ref{sec:discussion}, and will be utilized for the computation of the surface gravity in section \ref{sec:main}.

%%%%%%%%%%%%%%%%%%%%%%%%%%%%%%%%%%%%%%%%%%%%%
%%%%%%%%%%%%%%%%%%%%%%%%%%%%%%%%%%%%%%%%%%%%%

\section{Dressing as a passive supertranslation}
\label{sec:dressing}
%When studying spacetime deformations, one might wonder through what mechanisms would they affect the Hawking radiation.
In this section we will introduce the dressing process as a mechanism to affect the Hawking radiation%one, the dressing process \cite{Javadinazhed:2018mle}
, where the basis properly factorizing the Hilbert space of a scalar field in $\mathcal{I}^+$ described by the retarded version of eq.~\eqref{eq:gBMS}, i.e. the dressed states, are constructed and shown to be different from usual Fourier modes. However in the case of supertranslation the dressing process reduces to a phase shift, i.e. an active covariant transformation \cite{Javadinazhed:2018mle}, leaving the Hawking radiation unmodified.

Likewise in \cite{Bousso:2017dny} it is shown that the zero-frequency gauge particles (corresponding to the covariant transformation) are decoupled from the non-zero-frequency particles by the dressing factorization, rendering the BMS transformation and the associated \COMR{function }$f(\Theta)$ irrelevant to the black hole information paradox, invalidating the attempts in \cite{Hawking:2016msc,Hawking:2016sgy}.
We will discuss the issue in section \ref{sec:discussion}, \COMR{but in this section we}\NEW{and} mainly focus on generalizing and applying the dressing process to \COMR{the tBMS transformation}\NEW{tST} considered in section \ref{sec:tdtransform}, and deducing the corresponding modification to the Hawking radiation spectrum \NEW{in this section}.

\subsection{Dressed scalar fields near the horizon}

First noticed in \cite{Javadinazhed:2018mle} the dressing of a massless scalar field $\phi$ in $\mathcal{I}^+$ and the asymptotically Rindler region near $\mathcal{H}$ of a supertranslated Vaidya metric can be approximated by a time translation
\begin{align}
\hat\phi \left(v \right) = \phi \left(v - f\left(\Theta \right) \right) \;,\quad
\hat\phi \left(u \right) = \phi \left(u - f\left(\Theta \right) \right) \;,
\end{align}
where $v$ and $u$ are the advanced and retarded time, $\phi \left(v \right)$ and $\phi \left(u \right)$ are the incoming and outgoing modes, and $\hat\phi$ is the dressed scalar field. Such a translation can be considered as a covariant transformation of the scalar field by an supertranslation $-f$ that cancels out the passive supertranslation on the metric. The dressed field $\hat\phi$ would appear as if living on the vanilla Vaidya metric (c.f. \cite{Singh:2000sp}).
%
%To derive the Hawking radiation spectrum, the Bogoliubov transformation between the incoming and outgoing dressed states is constructed
%and therefore bear the same Bogoliubov transformation as a Schwarzschild metric with an identical mass at a certain moment $v$.

One may expand (un)dressed incoming/outgoing fields into Fourier modes $a_p^{}$, $\hat a_p^{}$, $b_p^{}$, $\hat b_p^{}$ as
\begin{align}
\phi \left(v \right) &= \int_0^\infty \frac{dp}{\sqrt{2\pi p}} \left( a_p^{} e^{-i p v} + a_p^\dagger e^{ i p v} \right)  \,,&  %\label{eq:phi_in}
\hat\phi \left(v \right) &= \int_0^\infty \frac{dp}{\sqrt{2\pi p}} \left( \hat a_p^{} e^{-i p v} + \hat a_p^\dagger e^{ i p v} \right)  \nonumber  \,,\\
\phi \left(u \right) &= \int_0^\infty \frac{dp}{\sqrt{2\pi p}} \left( b_p^{} e^{ i p u} + b_p^\dagger e^{-i p u} \right)  \,,& \label{eq:phi_out}
\hat\phi \left(u \right) &= \int_0^\infty \frac{dp}{\sqrt{2\pi p}} \left( \hat b_p^{} e^{ i p u} + \hat b_p^\dagger e^{-i p u} \right)  \,.
\end{align}
Then the relation between the incoming and outgoing modes, i.e. the Bogoliubov transformation is
\begin{align}
     b_p^{}      =\int_0^\infty dq\left(    \alpha_{pq}^{}     a_q^{}      +    \beta_{pq}^{}  a_q^\dagger\right)  \,,&\quad  \nonumber
\hat b_p^{}      =\int_0^\infty dq\left(\hat\alpha_{pq}^{}\hat a_q^{}      +\hat\beta_{pq}^{} \hat a_q^\dagger\right)  \,,\\
     b_p^\dagger =\int_0^\infty dq\left(    \alpha_{pq}^*      a_q^\dagger +    \beta_{pq}^*  a_q^{} \right)  \,,&\quad
\hat b_p^\dagger =\int_0^\infty dq\left(\hat\alpha_{pq}^* \hat a_q^\dagger +\hat\beta_{pq}^* \hat a_q^{} \right)  \,.  \label{eq:bogo}
\end{align}
Notice that the notation for the (un)dressed Bogoliubov coefficients is opposite of what is employed in \cite{Javadinazhed:2018mle}. Given the simple relation between $\phi$ and $\hat\phi$, Bogoliubov coefficients could be related by
\begin{align}
\hat\alpha_{pq}^{} \equiv \iint\! dk \!\; dl \!\; \tilde\alpha_{pk}^u \!\; \alpha_{kl}^{} \!\; \tilde\alpha_{ql}^v      = e^{i\left(p+q\right)f} \alpha_{pq}^{}  \:\,,\;\;\;
\hat \beta_{pq}^{} \equiv \iint\! dk \!\; dl \!\; \tilde\alpha_{pk}^u \!\;  \beta_{kl}^{} \!\; \tilde\alpha_{ql}^{v\,*} = e^{i\left(p-q\right)f}  \beta_{pq}^{}  \:\,,  \label{eq:bogo_dressed}
\end{align}
where $\tilde\alpha^u$ and $\tilde\alpha^v$ are the Bogoliubov coefficients of the dressing procedure at $\mathcal{I}^+$ and $\mathcal{H}$ respectively. We omit $\tilde\beta$ as we only consider coordinate transformations that preserves the covering of the coordinates. The bi-spectrum of outgoing modes $S_{pq}$ could be expressed as
\begin{align}
\mathcal{S}_{pq} \equiv \big\langle b_p^\dagger b_q^{} \big\rangle = e^{-i \left(p - q \right) f} \big\langle \hat b_p^\dagger \hat b_q^{} \big\rangle  \:\,.
\end{align}
Clearly the dressing only induces a phase shift factor%In can be seen that the dressing produces a phase factor in the flux of outgoing particles as compared with the derivation of Hawking.
, which is rendered $1$ in the case of the Hawking radiation as the bi-spectrum of the Vaidya metric contains a delta function $\delta \left( p-q \right)$. Actually by the same argument it wouldn't appear in any $n$-spectrum, leaving the Hawking radiation completely unmodified. Obviously we need a time-dependent transformation to generate a time-dependent phase shift for a non-trivial Bogoliubov transformation.
\subsection{A generic dressing procedure}
\label{sec:bogo}
Following the procedure in \cite{Javadinazhed:2018mle}, we formalize the dressing due to a generic coordinate transformation as a Bogoliubov transformation between the transformed proper basis (dressed) and the not-yet-transformed improper basis (undressed), and study its effect on the Noether currents.

Let us consider a Lagrangian $L$ for a field $\chi$, and its Noether currents (such as $T_{\mu\nu}$)
\begin{align}
j^\mu_{\,M} = \frac{1}{\sqrt{-g}}\lambda^\mu_{\,M} - \frac{\delta L}{\delta \nabla_\mu \chi} \delta_M \chi \:\,,
\end{align}
where $g \equiv \det g_{\mu\nu}$, $\delta_M$ is the variation against $M$-th generator, and $\partial_\mu \lambda^\mu_{\,M} = \delta_M \left( \sqrt{-g} L\right)$ is the boundary term (will be neglected for brevity). Under a generic coordinate transformation $\hat X^\mu \left( X^\nu \right)$ and assuming the generators transform as a type $(p,q)$ tensor, $j^\mu_{\,M}$ transforms contravariantly as
\begin{align}
\hat j^\mu_{\,M} = \frac{\partial \hat X^\mu}{\partial X^\nu} \left( \bigg( \underset{i=1}{\overset{p}{\otimes}} \, \frac{\partial \hat X}{\partial X} \bigg)
\bigg( \underset{k=1}{\overset{q}{\otimes}} \, \frac{\partial X}{\partial \hat X} \bigg) \right)_{\!M}^{\;N} j^\nu_{\,N}
\equiv \scJmunu \SMN j^\nu_{\,N} \:\,,
\end{align}
where $\mathcal{J}$ is the Jacobian from $X$ to $\hat X$, $\otimes$ is the tensor product, $S_M^{\:N}$ is the structure function of the generators, and the indices for the (contra-)co-variant transformation of the generators are omitted for brevity. Also from now on operators with or without hats are operators in $\hat X$ or $X$ coordinates respectively. Assuming the existence of orthonormal basis $\hat a_\mathbf{p}^{}$ covering the Hilbert space, with the parameters $\mathbf{p}$ forming the generators of $\hat X^\mu$ as $\hat \nabla_\mu \hat a_\mathbf{p}^{} \equiv \hat{\mathcal{F}}_\mu \big[ \mathbf{p}\big] \hat a_\mathbf{p}^{}$,
%\begin{align}
%\frac{\partial}{\partial \hat X^\mu} \hat a_\mathbf{p}^{} = \hat{\mathcal{F}}_\mu \big[ \mathbf{p}\big] \hat a_\mathbf{p}^{} \;\,,
%\end{align}
the coordinate transformation could be recast in the orthonormal form
\begin{align}
\hat j^\mu_{\,M} \big[ \hat a_\mathbf{p} \big]
\equiv - \frac{\delta L}{\delta \hat a_\mathbf{p}^{}} \frac{\delta_M \hat a_\mathbf{p}^{}}{\hat{\mathcal{F}}_\mu}
= \scJmunu \SMN j^\nu_{\,N} \big[ \hat a_\mathbf{p}^{} \big]
=-\scJmunu \SMN \frac{\delta L}{\delta \hat a_\mathbf{p}^{}} \frac{\delta_N \hat a_\mathbf{p}^{}}{\nohatscF_\nu}  \:\,,\quad
\scJmunu \underset{\hat a_\mathbf{p}^{}}{\sim}
\frac{\nohatscF_\nu \big[ \mathbf{p} \big]}{ \hat{\mathcal{F}}_\mu \big[ \mathbf{p} \big]}  \:\,,
\end{align}
where $\hat{\mathcal{F}}_\mu \big[ \mathbf{p}\big] \hat a_\mathbf{p}^{} \equiv \hat \nabla_\mu \hat a_\mathbf{p}^{}$. However if carelessly re-purposing $\hat a_p^{}$ for the $X$ coordinates without modifying the generating structure (denoted as $a_p^{}$, i.e. undressed modes), one would find
\begin{align}
\frac{\delta L}{\delta \hat a_\mathbf{p}^{}} \frac{\delta_M \hat a_\mathbf{p}^{}}{\hat{\mathcal{F}}_\mu}
&= \scJmunu \SMN \frac{\delta L}{\delta \hat a_\mathbf{p}^{}} \frac{\delta_N \hat a_\mathbf{p}^{}}{\nohatscF_\nu} 
 = \scJmunu \SMN \frac{\delta L}{\delta a_\mathbf{p}^{}} \frac{\delta_N a_\mathbf{p}^{}}{\hat{\mathcal{F}}_\nu}  \:\,,\quad
j^\mu_{\,M} \big[ \hat a_\mathbf{p} \big] =
\scJmunu \SMN j^\nu_{\,N} \big[ a_\mathbf{p} \big]  \:\,,\\
\hat a_\mathbf{p}^{} &=
\exp \left( \hat{\mathcal{F}}_\nu \big[ \mathbf{p} \big] \left( \frac{\partial}{\partial \hat{\mathcal{F}}_\nu} - \frac{\partial}{\partial \nohatscF_\nu} \right) \left[ \frac{\partial}{\partial \mathbf{p}} \right] \right) a_\mathbf{p}^{}  \;\,,  \label{eq:dressing}
\end{align}
where $\partial / \partial\mathcal{F}_\nu$ and $\partial / \partial\hat{\mathcal{F}}_\nu$ are the annihilator of $\mathcal{F}_\nu$ and $\hat{\mathcal{F}}_\nu$ respectively, i.e. $X^\nu$ and $\hat X^\nu$.

Clearly the dressing procedure has two impacts on the observables. First, the dressing induces a coordinate transformation, ensuring the general covariance at the macroscopic scale. Second, the same coordinate transformation would alter the orthonormal basis in a general covariant way, leading to a potentially non-trivial relation between the dressed and undressed states.

Such a relation could be expressed using the Bogoliubov transformation
\begin{align}
\hat a_\mathbf{p}^{} &= \int d\mathbf{q} \left( \int d\mathbf{X} \, e^{i \mathbf{p} \,\cdot\, \mathcal{F}^{-1} \big[ \hat{\mathbf{X}} \big]  - i \mathbf{q} \,\cdot\, \mathcal{F}^{-1} \big[ \mathbf{X} \big] } a_\mathbf{q}^{} + \int d\mathbf{X} \,
e^{i \mathbf{p} \,\cdot\, \mathcal{F}^{-1} \big[ \hat{\mathbf{X}} \big]  + i \mathbf{q} \,\cdot\, \mathcal{F}^{-1} \big[ \mathbf{X} \big] } a_\mathbf{q}^\dagger \right)  \,,
\end{align}
where $\mathcal{F}^{-1}$ is the inverse function of $\mathcal{F}$ and $A \cdot B = A_\mu B^\mu$ is the inner product.
Comparing with the Bogoliubov coefficient in eq.~\eqref{eq:bogo_dressed}, we can express $\tilde\alpha$ as
\begin{align}
\tilde\alpha_\mathbf{pq}^{} = \int d\mathbf{X} \, e^{i \mathbf{p} \,\cdot\, \mathcal{F}^{-1} \big( \hat{\mathbf{X}} \big)  - i \mathbf{q} \,\cdot\, \mathcal{F}^{-1} \big( \mathbf{X} \big) }  %\,,\quad
%\tilde \beta_\mathbf{pq} = \int d\mathbf{X} e^{i \mathbf{p} \,\cdot\, \mathcal{F}^{-1}\left( \hat{\mathbf{X}} \right)  + i \mathbf{q} \,\cdot\, \mathcal{F}^{-1}\left( \mathbf{X} \right) }
\:.
\end{align}

\subsection{time-dependent %\COMR{BMS transformation}\NEW{
supertranslation%}
case}
\label{sec:dressing_tST}

In this subsection we will derive the modification to the particle states and the associated Hawking radiation spectrum, due to \COMR{the tBMS transformation}\NEW{tST} introduced in section \ref{sec:tdtransform}.
For simplicity we only consider almost radially outgoing modes in the Eikonal limit with $\mathbf{q} = \omega du + Y_A d\Theta^A + O(f^2)$ where $Y_A = O \left(D_A f \right)$ is irrelevant to the derivation. Eq.~\eqref{eq:dressing} then reduces to up to $O(f)$
\begin{align}
:%\left( 1-f' \right)^{\frac{1+p}{2}}
e^{i \omega f} : \hat b_\omega^{} \approx b_\omega^{} \;\,,
\end{align}
where $:\quad:$ is the normal ordering operator, $b_\omega^{} \equiv b_\mathbf{q}^{}$ with $\mathbf{q}_u = \omega$, and $f \equiv f(\partial_\omega)$ is the abuse of notation.
Furthermore we utilize the Heisenberg picture to rewrite the relation as $: e^{i\omega f} : \hat b_\omega^{} = e^{ i \omega \tau} : e^{i\omega f_\tau} : e^{ -i \omega \tau} \hat b_\omega^{}$, where $f_\tau$ is $f$ after time translation by $\tau$.
%Furthermore we impose the adiabatic condition $ f''(u) \ll M^{-1}$ on the entire manifold, and utilize the Heisenberg picture to rewrite the relation as 
%\begin{align}
%: e^{i\omega f} : \hat b_\omega^{} = e^{ i \omega \tau} : e^{i\omega f_\tau} : e^{ -i \omega \tau} \hat b_\omega^{}  \;\,,
%\end{align}
%where $f_\tau$ is $f$ after time translation by $\tau$.
Now we may impose the adiabatic condition $ f''(u) \ll 2\pi kT \equiv \kappa$ % on the entire manifold 
where $kT$ is the Hawking temperature, and approximate $f_\tau$ by
%Given $\omega = O \left( M^{-1} \right)$ for the Hawking radiation and the adiabatic condition we have 
\begin{align}
f_\tau \left( x \right)
&\approx f_\tau^{(0)} + f_\tau^{(1)} x + \frac{1}{2!} f_\tau^{(2)} x^2 - \frac{2}{3!} f_\tau^{(2)} \kappa x^3 + \cdots
\approx f_\tau^{(0)} + f_\tau^{(1)} x + f_\tau^{(2)} \frac{1 - \left( 1+ \kappa x \right) e^{ -\kappa x }}{\kappa^2}  \nonumber  \\
&\approx f_\tau^{(0)} + f_\tau^{(1)} x + \frac{ e^{ \left( -\kappa + f_\tau^{(2)} \right) x } - 1 }{ - \kappa + f_\tau^{(2)} } - \frac{e^{ -\kappa x } - 1 }{-\kappa}  \:\,,  \label{eq:fform}
\end{align}
%and $f'_\tau = f_\tau^{(1)}$ 
where $f_\tau^{(n)}$ are the coefficients of this specific form of $f$ at $u = \tau$, that will be made clear in section \ref{sec:discussion}. Apparently the first and second terms correspond to a phase shift and a momentum rescaling respectively, while the rest leads to an Unruh effect that modifies the Hawking temperature. Plugging back into the relation we have
\begin{align}
b_\omega^{} &\approx %\;%\left( 1 - f_\tau^{(1)} \right)^{\frac{1+p}{2}}
%: e^{i\omega \left( f_\tau^{(0)} + f_\tau^{(1)} \left( - i \partial_\omega - \tau \right) \right) + \left( e^{ \left( T +f_\tau^{(2)} \right) \left( - i \partial_\omega - \tau \right)} -1 \right) / \left( T +f_\tau^{(2)} \right) - \left( e^{-i T \partial_\omega} - 1 \right) / T} : \hat b_\omega^{} \nonumber\\
%&= %\left( 1 - f_\tau^{(1)} \right)^{\frac{1+p}{2}}
e^{i\omega \left(  f_\tau^{(0)} - f_\tau^{(1)} \tau \right)} %\nonumber\\
%& \times
\hat b_{\left(1+f_\tau^{(1)}\right)\omega}^{-f_\tau^{(2)}} \equiv %\left( 1 - f' \right)^{\frac{1+p}{2}}
e^{i\omega \left(  f - f' u \right)} \hat b_{\left(1+f'\right)\omega}^{-f''}  \:\,,
\end{align}
where the superscript above $b$ is the modification to $\kappa_H$. The last equality is yet another abuse of notation as $\tau$ is set to be $u$, and $f$, $f'$ and $f''$ are now $f_\tau^{(0)}$, $f_\tau^{(1)}$ and $f_\tau^{(2)}$ respectively where the meaning of ``$f$'' will be clear in section \ref{sec:discussion}.
The effect on incoming modes $a_\mathbf{p}^{}$ can be carried out similarly by inversing the relation.
We then obtain the modification to the Hawking spectrum
\begin{align}
\big\langle g\left(q\right) {b^\dagger}_q {b^{}}_q \big\rangle &\approx
\left( 1+f'\right)^2 \big\langle g\left(q\right) \hat {b^\dagger}_{\left(1+f'\right)q}^{-f''} \hat {b^{}}_{\left(1+f'\right)q}^{-f''} \big\rangle
=
\left( 1+f'\right)^2 \big\langle g\left(q\right) \hat\beta_{\left(1+f'\right)qp}^{* \, \kappa \rightarrow \kappa - f''} \hat\beta_{\left(1+f'\right)qp}^{\kappa \rightarrow \kappa - f''} \hat a_p^{} \hat a_p^\dagger \big\rangle  \nonumber  \\
%&\approx \left( 1+f'\right)^2 \big\langle g\left(q\right) \hat\beta_{\left(1+f'\right)qp}^* \hat\beta_{\left(1+f'\right)qp}^{} \big\rangle
&= \big\langle g\left(q\right) \hat b_q^\dagger \hat b_q^{} \big\rangle \Big|_{ q \rightarrow \left( 1+f'\right) q  \,,\;  \kappa \rightarrow \kappa - f''}
\approx \big\langle g\left(q\right) \hat b_q^\dagger \hat b_q^{} \big\rangle \Big|_{\kappa \rightarrow \kappa - \kappa f' - f''}  \;,  \label{eq:dressing_tST}
\end{align}
where $g(q)$ is the density of states at energy $q$. The effect thus is equivalent to the Doppler shift and the Unruh effect due to the motion of the future null asymptotic observer. %relative to the black hole horizon. This motion, according to eq.~\eqref{eq:dressing_tST}, should be $\partial_u f$.
The physicality of such a motion is verified by the proper acceleration felt by a incoming particle along $dv$ direction, which happens to be $\partial_u \partial_u f$, i.e. the amount needed to explain the Unruh effect.

\NEW{Let us now clarify the assumptions made (Eikonal limit, adiabatic condition $f''(u) \ll \kappa$, and the specific form of $f$ as described by eq.~\eqref{eq:fform}) and thus the setup chosen in this subsection. First, the Eikonal limit we take guarantees the ``quasi-locality'' of the modes, which in conjunction with the other assumptions leads to a nearly Rindler patch at $\mathcal{I}$ that leads to eq.~\eqref{eq:dressing_tST} \cite{PhysRevD.83.041501}. The global properties of $\mathcal{I}$, however, can be very different from that of Rindler (e.g. tSTed or FSTed Vaidya considered in section \ref{sec:main}). Secondly, it is now clear that eq.~\eqref{eq:fform} ensures the thermality of the spectrum, without which the two point correlation function could be non-thermal and non-local \cite{1981PhRvD..24.2100S,Hsiang:2019dyp}, and would be too complicated for us to analyze. On the other hand given the adiabatic condition $f''(u) \ll \kappa$, in principle an analytic formula for the two point correlation function can be derived by the stochastic field approximation \cite{Raval:1996vt}. However, we will leave it as a future work given its irrelevance within the context of this work.}

\COMR{Since according to eq.~\eqref{eq:responseTf} $\partial_u f$ is related to the energy momentum tensor, it is natural to ask whether $f'$ in eq.~\eqref{eq:responseTf}, i.e. in \COMR{tST}\NEW{FST} (active) transformation is related to $f'$ in \COMR{tBMS}\NEW{tST} (passive) transformation. We will discuss the relation between two transformations in section \ref{sec:discussion}. N}\NEW{Finally, n}otice that the dressing of the incoming modes, i.e. $\partial_v f$ is irrelevant to the Hawking radiation. Therefore the associated $T_{vv}$ couldn't be memorized by the Hawking radiation through the dressing. This part of the information thus requires another channel to register, which we will discuss in the next section.

%\COM{The observation of the correlation function has been carried out in the recent experiment of analog black hole \cite{article}. We expect that the modification of correlation function \ref{eq:dressing_\COMR{tBMS}\NEW{tST}} would appear in the physical observable.}\HWC{This paragraph seems weak. Better not presented than weak.}
%In a physical point of view, the dressing as a passive BMS transformation is just a coordinate transformation that it can always be shifted into a physical equivalent vacuum. Consequently, the effect is coordinate dependent as our expectation.

%%%%%%%%%%%%%%%%%%%%%%%%%%%%%%%%%%%%%%%%%%%%%
%%%%%%%%%%%%%%%%%%%%%%%%%%%%%%%%%%%%%%%%%%%%%

\section{Hawking radiation of dynamical black hole}
\label{sec:main}
Another important aspect of the Hawking radiation is the horizon and its associated surface gravity.
In this section we first adopt the ray-tracing method presented in \cite{PhysRevD.83.041501} and derive the relation between the Hawking temperature and the in-affine surface gravity for a generic spacetime.
Then we demonstrate that from the null foliation point of view, \COMR{the tST}\NEW{FST} and the associated \COMR{tBMS transformations}\NEW{tST} are completely indistinguishable after applying the 1st order approximation in $f$. To wit, \COMR{tST}\NEW{FST} (including SST case in \cite{Chu:2018tzu})\COMR{ transformation} merely induces a spurious effect on the surface gravity, a byproduct of the covariant transformation.

However we realized that unlike the original Vaidya black hole, for the \COMR{tST}\NEW{FST}ed one the infinite redshift surface does not coincide with the apparent horizon%, leading to the development of an ergosphere
. The discovery indicates that the singular structure of the null foliation is not simple and would be smeared by a n\"{a}ive %ly applying 
1st order approximation in $f$. To eradicate this issue we construct an exact null foliation, carefully apply the approximation, and obtain the correct location of the apparent horizon and its associated surface gravity up to 1st order in $f$. The physical meaning of the %indistinguishabitlity between \COMR{tST}\NEW{FST} and \COMR{tBMS}\NEW{tST} transformations
exact foliation, the %newly found ergosphere, 
corrected apparent horizon and its surface gravity will be discussed in section \ref{sec:discussion}.

%In \cite{Chu:2018tzu}, the surface gravity of a Vaidya black hole under BMS transformation has been studied and we will briefly review the result in the following subsections. For a dynamical spacetime, the spacetime is not transformed into the original one during the accretion process of the black hole. Thus, no explicit Killing vector could be defined to determine the surface gravity. We have to find an appropriate definition of the null vector of in-coming and out-going rays. 

\subsection{Ray-tracing, surface gravity, and the covariance of the Hawking temperature}
\label{sec:HawkingTemp}

As shown in \cite{PhysRevD.83.041501}, given a double null foliation with the advanced and retarded time $(v,u)$ \cite{Huber:2019yfg}, we can construct a ray-tracing function $v = p(u)$ to mark the center of the foliation %(a geodesic $\Gamma$ at areal radius $r \to 0$) 
where incoming rays from $\mathcal{I}^-$ labelled $v$ ``reflect'' %and become outgoing rays 
toward $\mathcal{I}^+$ labelled $u$, and the Hawking temperature $\kappa_H$ under the adiabatic condition $\left| \dot \kappa_H \right| \ll \kappa_H^2$ ($\dot F \equiv dF/du$) in natural units $h=G=1$ becomes
\begin{align}
2\pi kT \equiv \kappa_H = - \left( \ln \dot p \right) \hspace{-3.3pt}\dot{\color{white}p}  \,.
\end{align}

\begin{figure}[!ht]
\centering
\vspace{-3pt}
\includegraphics[width = .6\textwidth]{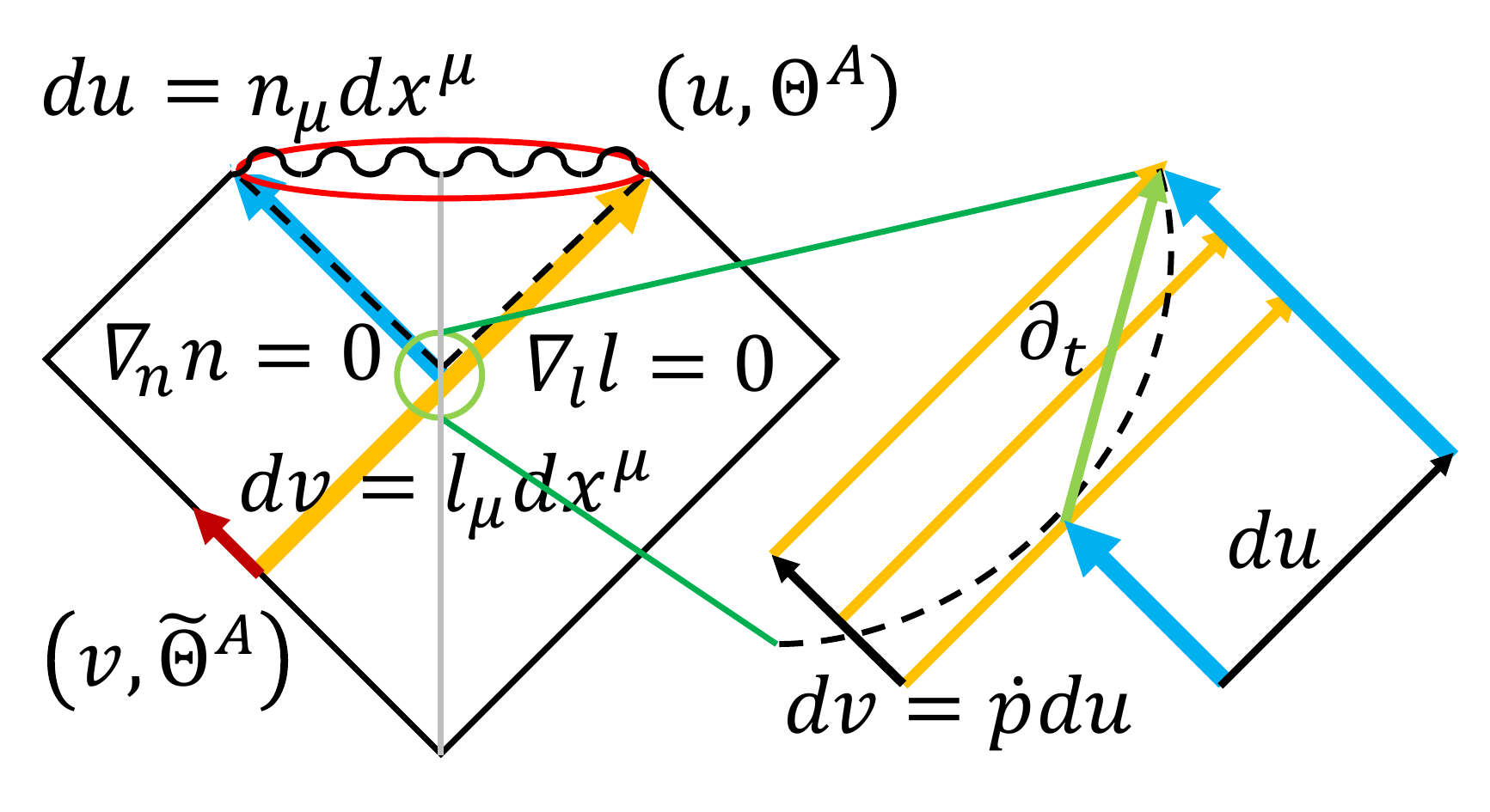}
\vspace{-10pt}
\caption{Penrose-like diagram for the double null foliation $(u,v)$ %utilized in this subsection
, with left and right parts representing hypersurfaces along incoming and outgoing directions respectively. The blue and yellow arrows represent the congruence of $u$ and $v$, while the green arrow being the gradient along the tangent vector $\partial_t$ underlying the ray tracing function $v=p(u)$. Notice that $du$ is already living on the same slice as $dv$.}\label{fig:BH}
\end{figure}

We have to emphasize that the ray-tracing function is actually an abuse of notion by virtue of ignoring the ray direction. With isotropy, it is fine to focus only on the quotient space of $S^2$, but without isotropy, it is exceedingly dangerous as $v$ in the ray-tracing function refers to $(v,\tilde \Theta^A)$ rather than to the advance time $(v,\Theta^A)$ associated with $u$, where $\Theta^A$ and $\tilde \Theta^A$ are respectively the outgoing and the incoming directions. To circumstance this issue we choose $ g_{uA} = 0$ and transport $du$ from $\Theta^A$ to $\tilde \Theta^A$, allowing us to properly construct the level set $v - p \left( u \right)$ and its tangent vector $\frac{\partial}{\partial t} \propto \frac{\partial}{\partial u} + \dot p \, \frac{\partial}{\partial v}$ embodying the ray-tracing function, as shown in figure \ref{fig:BH}. Notice that the other choice, i.e. transporting $dv$, could lead to outgoing rays trapped by the black hole, thus unfavorable.

For the sake of simplicity we will introduce the null geodesic congruence $n$, $l$ as
%Now we may relate $\kappa_H$ to the surface gravity by introducing the null geodesic congruence $n$, $l$ as
\begin{align}
l_\mu \equiv \nabla_\mu v  \:\,,\quad
n_\mu \equiv \nabla_\mu u  \:\,,\quad
%n^A \partial_A = 0  \:\,,\quad
n_\mu l^\mu \equiv -2 \Omega^{-1} \;,\quad
\dot p = \frac{t^\mu \nabla_\mu v}{t^\mu \nabla_\mu u} \;\rightarrow\;%,\quad
t^\mu \equiv \Upsilon \left( l^\mu + \dot p \, n^\mu \right)  \:,%\quad
%\nabla_l l^\mu = \nabla_n n^\mu = \nabla_t t^\mu = 0  \;\,,
\end{align}
where $\Omega$ is the redshift, $\Upsilon$ is the time dilation factor, and the Latin indices are raised or lowered by $g_{\mu\nu}$.
The Hawking temperature thus is related to a trajectory (assumed to be geodesic) describing the event horizon of a classical black hole , falling materials in the no-horizon proposal, etc.
From the geodesic equation $\nabla_t t^\mu = 0$ where $\nabla_X \equiv X^\mu \nabla_\mu$% is the directional derivative
, we have
\begin{align}
0 &= \Upsilon^{-1} \nabla_t \left( \Upsilon l^\mu + \Upsilon \dot p \, n^\mu \right) = \left( \nabla_t \ln\Upsilon \right) l^\mu + \nabla_t \ln \left( \Upsilon \dot p \right) \dot p \, n^\mu + \Upsilon \dot p \left( \nabla_n l^\mu + \nabla_l n^\mu \right)  \nonumber\\
&= \nabla_t \ln\Upsilon \, l^\mu + \nabla_t \ln \left( \Upsilon \dot p \right) \dot p \, n^\mu - \Upsilon \dot p \left( \nabla_n \ln\Omega \, l^\mu + \nabla_l \ln\Omega \, n^\mu \right) + S^2\text{ terms}  \nonumber  \;\,,\\
&\quad\!\;\: \nabla_t \ln\Upsilon = \Upsilon \dot p \, \nabla_n \ln\Omega  \;\,,\quad\:
\nabla_t \ln \left( \Upsilon \dot p \right) = \Upsilon \nabla_l \ln\Omega  \;\,,
\end{align}
where $d\Theta^A$ terms are neglected for simplicity. Then the Hawking temperature is translated as
\begin{align}
\kappa_H = -\frac{\nabla_t \ln \dot p}{\nabla_t u} = \frac{- \Upsilon \left( \nabla_l \ln\Omega - \dot p \, \nabla_n \ln\Omega \right)}{-2 \Omega^{-1} \Upsilon} &= \kappa_L - \dot p \, \kappa_N \:\,,\\
\nabla_{\left( \Omega l \right)} \left( \Omega l_\mu \right) \equiv 2\kappa_L \Omega l_\mu  \;,\quad
\nabla_{\left( \Omega n \right)} \left( \Omega n_\mu \right) &\equiv 2\kappa_N \Omega n_\mu  \;,
\end{align}
where $\kappa_L$ is the familiar in-affine surface gravity of the in-affine null geodesic 1-form
\begin{align}
\kappa_L = -\frac{1}{4} n^\mu \nabla_L L_\mu  \;,\quad
L \equiv \Omega l  \:\,,\quad
L_\mu n^\mu = -2  \;\,.
\end{align}
Interestingly we arrive at a form compatible with the first law of black hole thermodynamics \cite{PhysRevLett.92.011102}.

%Notice that even though the above expression of the Hawking temperature appears local, it actually stems from asymptotic variables $u$ and $v$ in the Bogoliubov transformation. The existence of both incoming and outgoing light rays from and to the null asymptotes is essential for the derivation of the Hawking radiation, indicating that certain globally defined horizon akin to the event horizon would be the boundary of the ``black hole photosphere''.

%The precise location of such a boundary is beyond the scope of this work, but with the adiabatic condition naturally one would assume that the apparent horizon should approach the boundary well enough. Thus from now on we would consider the apparent horizon as the emitting surface.

Surprisingly the Hawking temperature originated from the globally defined Bogoliubov transformation could be recast into a local form without relying on any globally defined object such as the Killing horizon or the event horizon, indicating that the emission itself is a local event. At every point multiple ray-tracing functions exist, each relating one outgoing ray $u$ to its associated incoming ray $v$. The perceived Hawking radiation thus is the integral effect along the line of sight, aggregating different rays ``reflected'' at different locations with varying temperatures. The observed radiation temperature then should be the average energy of observed particles.

Such an integration would be an interesting topic that deserves further investigation. However for simplicity we would approximate it by the dominating part, i.e. a region with the highest surface gravity. With the adiabatic condition the apparent horizon should be a reasonable choice, and from now on would be considered as the emitting surface. More thorough discussion about the physical implication of \cite{PhysRevD.83.041501} and the choice of the emitting surface will be presented in section \ref{sec:discussion}.

The apparent horizon, a.k.a. the marginally trapped closed surface, is a closed surface that separates the spacetime along the line of sight into a trapped region where classically no light can escape and another where light may escape should it never cross the horizon again. It is defined by the congruence $l$, $n$ and their associated expansion rates of the surface area element $\theta_L$, $\theta_N$ as
\begin{align}
%h_{\mu\nu} &= g_{\mu\nu} - \frac{l_\mu n_\nu + n_\mu l_\nu}{l_\sigma n^\sigma} \;,\\
\theta_L = \left( l_\sigma n^\sigma \right)^{-1} \nabla_\mu l^\mu = \nabla_\mu L^\mu - 2 \kappa_L = 0  \:\,,\quad
\theta_N = \left( l_\sigma n^\sigma \right)^{-1} \nabla_\mu n^\mu < 0  \:\,.
\end{align}
%where $X = L \lor n$, the apparent horizon is formally defined as the outermost closed surface with a vanishing $\theta_L$ and a negative $\theta_N$.
%Assuming the null energy condition we may
Finally given $\dot p \propto e^{- \int \kappa_H du}$ a horizon would quickly approach nullity after its formation, leading to $\kappa_H \approx \kappa_L \equiv \kappa$. Our goal, i.e. to find out the Hawking temperature of a \COMR{tST}\NEW{FST}ed or a \COMR{tBMS}\NEW{tST}ed black hole, thus reduces to the setup of the 1-form $L$ and the identification of the apparent horizon.

Obviously both the surface gravity and the apparent horizon depend only on the observer's line of sight, and transform covariantly under coordinate transformation. % We expect that the form of surface gravity remains unchanged after having a coordinate transformation on the metric
%\begin{align}
%     g_{\mu\nu} \rightarrow g'_{\mu\nu} = \mathcal{L_{\zeta}}g_{\mu\nu} \quad , \kappa \rightarrow \kappa .
%\end{align}
For example, \COMR{the tBMS transformation}\NEW{tST} would act on the surface gravity at the apparent horizon of a Vaidya black hole as
\begin{align}
\kappa^{\!\text{Vaidya}}
&=\kappa_L \left( v, r = r_H \left( v \right) , \Theta^A \right)
= \frac{1}{ 4 M \left( v \right) }  \nonumber\\
&\rightarrow \kappa^\text{\COMR{tBMS}\NEW{tST}}
= \frac{1}{ 4 M \left( v + f \left( v, \Theta^A \right) \right) } 
\approx \frac{1}{ 4 M \left( v \right) } \left( 1- \frac{M'\left( v\right)}{M\left( v\right)} f \left( v, \Theta^A \right) \right)  \,,  \label{eq:kappa_tST}
\end{align}
where $r_H$ is the radius of the apparent horizon.
%For surface gravity, we identify that the \COMR{tBMS}\NEW{tST} transformation acts as a linear transformation of scalar.
Interesting the surface gravity after \COMR{BMS transformation}\NEW{supertranslation} coincides with that of \cite{Chu:2018tzu}, suggesting that maybe we can interpret the effect of \COMR{tST transformation}\NEW{FST} on the surface gravity as an active covariant transform. If so \COMR{the tST transformation}\NEW{FST} would be rendered useless for the resolution of the information loss paradox, as least in the case of the surface gravity. We will falsify the above statement in the following subsection.

\subsection{%Apparent horizon and infinite redshift surface: 
Perturbative analysis of the %\COMR{time-dependent}\NEW{
flow-induced%}
supertranslated spacetime}
\label{sec:FST_perturb}

In this subsection we will test whether the surface gravity on the apparent horizon of a \COMR{tST}\NEW{FST}ed Vaidya black hole truly has the same form as that of a \COMR{tBMS}\NEW{tST}ed Vaidya black hole. Since \COMR{tST transformation}\NEW{FST} can be interpreted as a metric deformation procedure as depicted in eq.~\eqref{eq:FST}, we would expect the congruence $L$ and $n$ that foliates the spacetime into 2-spheres %with induced metric $h$ 
to acquire a similar deformation
\begin{align}
%g_{\mu\nu} &\rightarrow g_{\mu\nu}^\text{\COMR{tST}\NEW{FST}} = g_{\mu\nu} + \mathcal{L}_f g_{\mu\nu} \Big{|}_{f \rightarrow f\left(v\right)}  \;,\\
L_\mu \rightarrow L_\mu + \mathcal{L}_f L_\mu \Big{|}_{f \rightarrow f\left(v\right)} + O ( f^2 )  \:\,,\quad
n_\mu \rightarrow n_\mu + \mathcal{L}_f n_\mu \Big{|}_{f \rightarrow f\left(v\right)} + O ( f^2 )   \:\,.  \label{eq:lndeform}%\\
%\kappa &\rightarrow \kappa'  \;\,,  \label{eq:g'}
\end{align}
Indeed they are null geodesic up to $O(f)$. Furthermore the angular components can be extended by additional terms of $O(f)$ constant along the geodesic at the leading order, denoted as $\Delta L_A$ and $\Delta n_A$. These two terms correspond to the choice of the line of sight, and are necessary as the normal forms on the apparent horizon of the \COMR{tST}\NEW{FST}ed Vaidya black hole, contrary to that of the Vaidya black hole, may not be completely radial. An additional scaling $n \rightarrow \eta n \,,\; L \rightarrow L / \eta$ is also mandatory for $n$ to be a congruence, i.e. $dn = 0$, but given that $\eta$ only affects the surface gravity as $\kappa_L \rightarrow \kappa_L / \eta + \nabla_L \eta^{-1} / 2$ we may ignore it for now.
The resulting 1-forms up to $O(f)$ become
\begin{align}
n^\text{\COMR{tST}\NEW{FST}}_v &\approx -1  \:\,,\;&
n^\text{\COMR{tST}\NEW{FST}}_r &\approx  0  \:\,,\;&
n^\text{\COMR{tST}\NEW{FST}}_A &\approx -\scD_A f - \Delta n_A  \;\,,\\
L^\text{\COMR{tST}\NEW{FST}}_v &\approx -V + \frac{2M' f}{r} + \frac{ M \scD^2 f}{r^2}  \:\,,\;&
L^\text{\COMR{tST}\NEW{FST}}_r &\approx  2  \;\,,\;&
L^\text{\COMR{tST}\NEW{FST}}_A &\approx -\scD_A \left( \scD^2 f +V f \right) + \Delta L_A  \;\,.
\end{align}
The associated expansion rate, the horizon radius, and the surface gravity on the horizon are
\begin{align}
\theta_L^\text{\COMR{tST}\NEW{FST}} &= \frac{2}{r} \left( V - \frac{2M' f}{r} + \left( 2V -1 \right) \frac{\scD^2 f}{2r} + \frac{\scD^A \Delta L_A}{2r} \right) + O(f^2) \:\,,\\
r_H^\text{\COMR{tST}\NEW{FST}} &\approx 2M + \frac{1}{2} \left( 4 M'f + \scD^2 f - \scD^A \Delta L_A \right)  \;,\quad
\kappa_H^\text{\COMR{tST}\NEW{FST}} \approx \frac{1}{4M} \left( 1 -\frac{M'}{M} f + \frac{\scD^A \Delta L_A}{2M} \right)  \,,  \label{eq:kappa_FST}
\end{align}
where $r_H$ and $\kappa_H$ are derived after dropping terms of $O(f^2)$. As expected $\Delta L_A$ is directly related to the horizon deformation and the modification to the surface gravity. To determine $\Delta L_A$ and correspondingly $\kappa_H$, we need the constraint that $L$ and $n$ are the normal 1-forms of the horizon:
\begin{align}
\nabla_A \theta_L = \frac{-1}{2} \left(L_A \nabla_n + n_A \nabla_L \right) \theta_L  \;\,.  \label{eq:normal}
\end{align}
Up to $O(f)$ the above relation can be written as a 1st order differential equation in $\Delta L_A$ without the presence of $f$. This $f$-independency suggests that the outgoing ray from the horizon of \COMR{tST}\NEW{FST}ed black hole remains radial, i.e. $\Delta L_A = 0$. This is a manifestation of the fact that the tidal Love number of a 4$-$d black hole is identically zero.

Now we may turn back to the factor $\eta$ dropped before. By solving $d\left( n^\text{\COMR{tST}\NEW{FST}} \eta \right) =0$ we have $\eta \approx 1 + f'$. Consequently the perturbed surface gravity %as depicted by eq.~\eqref{eq:kappa_\COMR{tST}\NEW{FST}} reduces to
$\kappa_H = \frac{1}{4M} \left( 1 -\frac{M'}{M} f - f' \right) -f''$ where first two terms resembles that of eq.~\eqref{eq:kappa_tST} and can be attributed to an active covariant transformation on $\mathcal{I}^-$, while the other two terms have the same form as the Doppler effect and the Unruh effect derived in section \ref{sec:dressing_tST}, and is related to another active covariant transformation on $\mathcal{I}^+$ due to the radial velocity of the observer, which in turn comes from the tie between the observer and the global clocks $(v,u)$ in section \ref{sec:HawkingTemp}. We will discuss the choice of foliation more thoroughly in section \ref{sec:discussion}.%On the other hand eq.~\eqref{eq:lndeform} assumes a stationary observer, and we will stick to this choice of foliation in the following.

Apparently these effects are all erasable by \COMR{tBMS transformations}\NEW{tST}, and thus can not be considered a probe to the matter flow. However as the result above is obtained after applying the 1st order approximation in $f$, it may be invalid near the horizon. To check it we introduce the infinite redshift surface $\Omega\to 0$. Should the infinite redshift surface deviates from the apparent horizon, the linear approximation of the expansion rate $\theta_L$ fails, as it would contain at least a node corresponding to $\Omega\to 0$ whose cancellation with a pole is not guaranteed. Assuming a constant $M$ for simplicity, the radius of the infinite redshift surface for the outgoing null geodesic with $\Delta L_A = 0$ is
\begin{align}
r \big{|}_{\Omega\to 0} = 2M + 2M' f + \frac{1}{2} \scD^2 f + \frac{1}{2} \int_v^\infty e^{-\frac{\tilde{v}-v}{4M}} \scD^2 f' \left( \tilde{v} \right) d\tilde{v} + O(f^2)  \;\,,
\end{align}
where $\tilde{v}$ is the dummy variable. Indeed the infinite redshift surface does not coincide with the apparent horizon, suggesting the failure of the linear approximation near the horizon. In the following we will tackle this issue by delaying the perturbative analysis for as long as possible.

\subsection{Horizon deformation on an exact supertranslated black hole}
\label{sec:exact}
As discussed before an accurate form of the null 1-form $L$ near the horizon is essential for the precise location of the horizon. However $L$ is inherently ambiguous as both \COMR{tBMS}\NEW{tST} and \COMR{tST transformations}\NEW{FST} are carried out by the Lie derivative, precluding possible higher order terms of $f$ from the metric. %Luckily the location of the horizon is insensitive to the higher order terms as long as the system contains only one horizon near $r=2M$, as will be proven in appendix \ref{sec:appendix}. Thus we may choose whatever coordinates system that eases the analysis the most, which happens to be the one with the linear response relation as shown in eq.~\eqref{eq:responseTf}. The resulting metric
Luckily as will be discussed in section \ref{sec:discussion}, the combined transformation of \COMR{tST}\NEW{FST} and \COMR{tBMS}\NEW{tST} as depicted in eq.~\eqref{eq:responseTf} happens to be the unique, exact generator of the BMS symmetry without gravitational waves or the outgoing flow. While the associated metric
\begin{align}
ds^2 &= g_{\mu\nu} + \mathcal{L}_f g_{\mu\nu} \Big{|}_{f \rightarrow f\left(v\right)\!} \!- \mathcal{L}_{f\left(v\right)} g_{\mu\nu} = - \left( \left( 1-\frac{2M}{r} \right) \left( 1 - 2f' \right) - \scD^2 f' \right) dv^2  \nonumber  \\
&+ 2\left(1-f'\right)dv dr - 4r\scD_A f' dv d\Theta^A + r^2 \gamma_{AB} d\Theta^A d\Theta^B  \,, \label{eq:gexact}
\end{align}
%has very special structures that will be discussed in section \ref{sec:discussion}
% While this is by no mean the unique manifestation of the \COMR{time-dependent}\NEW{flow-induced} supertranslation, others would consist of outgoing flows that
can be written in an elegant form
, %but 
what matters most is the exact form of the null normal 1-form:
\begin{align}
L_v = - \left( 1-\frac{2M}{r} \right) &\left( 1 - 2f' \right) + \scD^2 f' - \scD_A \left( f' + \psi /r \right) \scD^A \left( f' + \psi /r \right) \nonumber  \:\,,\\
L_r = 2 &\left( 1 - f'\right)  \;\,,  \qquad\qquad
L_A = 2\scD_A \psi  \label{eq:L_exact}  \;\,,
\end{align}
where $\psi \left( v,r,\Theta^B \right)$ is the lensing potential and $2\scD_A \psi$ substitutes $\Delta L_A$ without loss of generality. The overall scaling $1/\eta$ is dropped due to its indistinguishability from \COMR{a tBMS transformation}\NEW{tST} on $\mathcal{I}^+$. The associated expansion rate $\theta_L$ (omitted for brevity) is a Pad\'e series of order [4/2] in r multiplied by $r^{-3}$, with 4 zeros and 2 non-trivial poles where half of them are spurious at $2M r = %\sqrt[3]{-1}^{ (1 \lor 3 \lor 5) }
e^{(1 \lor 3 \lor 5)\pi i/3}
\scD_A \psi \scD^A \psi + O(f^3)$, while the other two zeros %(at $r_{0,\pm}$) 
and the pole respectively are at%(at $r_\infty$) with
\begin{align}
r_{0,\pm} &= 2M + \frac{1}{2} \left( 2M \scD^2 f' - \scD^2 \psi \pm \Delta \right) + O(f^2)  \:\,,\quad  \label{eq:r_H}
r_\infty = 2M + 2M \scD^2 f' + O(f^2)  \:\,,\\
%and they correspond to two apparent horizons where
\Delta^2 &\equiv \left( 2M \scD^2 f' + \scD^2 \psi \right)^2 + 4 \scD_A \psi \scD^B \left( \psi + 2M \scD^2 f' - 4M \psi' \right)  \:.
\end{align}
By the cosmic censorship conjecture, we expect that a pole (singularity) should be hidden behind a zero (apparent horizon), leading to a requirement of $O(f)$ that suggests a natural substitution
\begin{align}
\scD_A \psi \scD^B \left( \psi + 2M \scD^2 f' - 4M \psi' \right)
&\equiv \scD_A \psi \scD^B \alpha  \gtrsim 0  \label{eq:ccc}  \:\,,\\
%\end{align}
%that suggests the following substitution
%\begin{align}
4M \psi'  \equiv \psi + 2M \scD^2 f' - \alpha  \:\,,\quad
\Delta^2 &\equiv \left( 2M \scD^2 f' + \scD^2 \psi \right)^2 + 4 \scD_A \psi \scD^B \alpha  \label{eq:alpha}  \:\,.%,\quad
%\scD_A \psi \scD^A \alpha \geq 0  \:\,.
\end{align}
Surprisingly eq.~\eqref{eq:ccc} is actually sharp, i.e. the cosmic censorship conjecture is fulfilled. We will prove this statement by requiring $n_\mu$ to be exact on the horizon as it is foliated by $\theta_L$ and $v$.

From eq.~\eqref{eq:normal} we may obtain $n_A$ on the horizon up to $O(f)$ as
%The apparent horizons are closed 2-spheres that give $\alpha$ a further constraint. We can find the $l_\mu^{exact}$ on the apparent horizon by substituting the zeros into eq.\ref{exact} and then obtain $n_\mu$ from eq.~\eqref{eq:normal}. The angle components of null congruence $n_A$ on $S^2$ are used to study the closure for the apparent horizon. We derive $n_A$ up to $O\left(f\right)$ by assuming that all higher order terms are small.
\begin{align}
n_A|_\mathcal{H} = \frac{\nabla_\mu \theta_L \nabla^\mu \theta_L}{\left( \nabla_L \theta_L \right)^2} L_A &- \frac{2\nabla_A \theta_L}{\nabla_L \theta_L} = \frac{1}{2M'} \bigg( \scD_A \left( -2M\scD^2 f' + \scD^2 \psi - 2\psi \mp \Delta \right)  \nonumber  \\
&+ \frac{ 2\scD^2 \alpha \left( 2M \scD^2 f' + \scD^2 \psi \mp \Delta \right) + 2\scD_B \alpha \scD^B \left( \psi + \alpha \right)}{\pm \Delta \left( 2M \scD^2 f' - \scD^2 \psi \pm \Delta \right)} \scD_A \psi \bigg)  \,.
\end{align}
Notice that all components are evaluated exactly except for $\nabla_r \partial_A \psi$ which is integrable along geodesic only upto $O(f)$. Luckily the higher order terms does not affect $n_A$ at $O(f)$.

%If the apparent horizon is a closed 2-sphere, the angle component of null congruence $n_A$ must satisfy
Now we may check the close condition $dn_A|_\mathcal{H} = 0$ where $d$ is the exterior derivative on the horizon. As the exterior derivative commutes with the projection operator, the only solution at $O(f)$ is apparently $\alpha=0$.
The lensing potential on the horizon then can be solved by eq.~\eqref{eq:alpha} as
\begin{align}
\psi \left(v \right) = -\frac{1}{2} \int_v^\infty e^{-\int_v^{v_1} \frac{1}{4M \left( v_2 \right)} dv_2} \scD^2 f' \left( v_1\right) dv_1 + O(f^2) \:\,. \label{eq:psi}
\end{align}
Clearly the lensing potential depends on the exponential average of the metric function $f'$ directly related to the anisotropic part of the radial energy flow, with a decay time of $4M$ toward the future.

With $\alpha = 0$ the radius of the zeros in eq.~\eqref{eq:r_H} reduces to
\begin{align}
r_\mathcal{E} = 2M + 2M \scD^2   f' + O(f^2)  \:\,,\quad
r_\mathcal{H} = 2M - \scD^2 \psi + O(f^2)  \:\,.
\end{align}
Notice that with $\alpha = 0$ all three singular structures of $\theta_L$ are near $r = 2M$, and thus the black hole remains a 2-sphere. One of the zeros at $r_\mathcal{E}$ actually coincides with the pole at $r_\infty$ and degenerates into the ergosphere ($L_v \to 0$), leaving the other at $r_\mathcal{H}$ the apparent horizon.

Now we may turn to the scaling factor $\eta$. By solving the geodesic equation up to $O(f^2)$ and the close condition $dn=0$ up to $O(f)$, we have $\eta$ and the surface gravity respectively as
\begin{align}
\eta &= 1 - \chi' + O(f^2) = 1 - \frac{1}{M'} \left( \scD^2-1 \right) \psi' + O(f^2) \:\,,\quad
d\chi|_\mathcal{H} \equiv n_A  \:\,,\\
\kappa_\mathcal{H}
&= \frac{ 1 -2 f' - \scD_A \left( f'+ \psi / r \right) \scD^A \left( f'+ \psi / r \right) r/ M}{4M \eta \left( 1-f' \right) \left( \frac{r}{2M} \right)^2} - \frac{f''}{1-f'} - \frac{\nabla_L \eta}{2 \eta^2}  \nonumber\\
%= \frac{1}{4M} \left( 1 - \scD^2 f' + \scD^2 \psi \mp \Delta \right)
&= \frac{1 + \scD^2 \psi / M - f' + \chi'}{4M} - f'' + \chi'' + O(f^2)  \:\,.
\label{exsur}
\end{align}
Notice that any effort of compensating the temperature anisotropy by the dressing, thus hiding the information about the energy flow, would be futile as $\scD^2\psi$ depends on the flow differently from that of section \ref{sec:dressing_tST}, and thus unavoidably requires the observer to access the information. We thus conclude that the Hawking temperature is indeed modified by the energy flow on the horizon, and one may reconstruct the flow by recording the Hawking temperature at $\mathcal{I}^+$. We will more thoroughly discuss the implication of this discovery on the information loss paradox in section \ref{sec:discussion}.

%%%%%%%%%%%%%%%%%%%%%%%%%%%%%%%%%%%%%%%%%%%%
%%%%%%%%%%%%%%%%%%%%%%%%%%%%%%%%%%%%%%%%%%%%

\section{Discussion and future works}
\label{sec:discussion}

\paragraph{The response function and the conserved charge of the exact metric}
In section \ref{sec:tdtransform}, \COMR{the tBMS transformation}\NEW{tST} is introduced as a coordinate transformation that in conjunction with \COMR{the tST transformation}\NEW{FST} forms a linear response relation between the parameter $f$ and the energy momentum tensor $T$ (linearized conserved charge of BMS symmetry) at $O(f)$. More precisely \COMR{the tBMS ``transformation''}\NEW{tST} within this context is the unique gauge choice where \COMR{the tST transformation}\NEW{FST} as the generator of $T$ at non-zero frequency is integrable along advance time $dv$.

This statement is in fact accurate even for a large $f$, as the nonlinear terms of $T$ can be proven to be of the form $T_1 \left( f' \right) + \partial_t T_2 \left( f' \right)$, composed of only the local quantity $f'$ and a boundary term (presumably anisotropy self-energy). This would suggest the existence of a conserved charge along the orthogonal direction of $dv$. Indeed $\nabla^\mu \left( T_{\mu\nu} n_\text{r}^\nu \right) = 0$ where $n_\text{r}^\nu \partial_\nu = \partial_r$ is the null geodesic congruence of $v$. Since $n_\text{r}$ is the asymptotically Minkowski direction of the BMS metric introduced in subsection \ref{sec:BMS} and it is indeed an asymptotic Killing vector field, the combined transformation of \COMR{tST}\NEW{FST} and \COMR{tBMS}\NEW{tST} fulfils our initial intent, i.e. to generate the BMS charge dynamically, and perhaps minimally as the incoming gravitational wave vanishes and $\mathcal{I}^+$ remains vacuum.

Notice however that there are several caveats. First, we are not claiming the capability of generating large \COMR{BMS transformation}\NEW{supertranslation} globally for an arbitrary spacetime satisfying eq.~\eqref{eq:BMSgauge}% nor the uniqueness of the method (which will be paramount in the next subsection)
, but merely one possible way for the Vaidya spacetime. Second, neither \COMR{tST}\NEW{FST} nor \COMR{tBMS}\NEW{tST} is an accurate depiction of the combined transformation, and either of them may be inextensible to the horizon. In fact, we are only certain of the first order form of the combined transformation, as it is the unique way of writing down an integrable dynamical generator of the BMS charge. Finally, although \COMR{tBMS}\NEW{tST} should \COMR{not} be regarded as a gauge fixing of \COMR{the tST transformation}\NEW{FST} and even may be inextensible to the horizon, \COMR{tBMS}\NEW{tST} as a coordinate transformation is well-defined on the null asymptotic region. What is presented in section \ref{sec:dressing_tST} remains valid for the asymptotic observer. Notice that while in \cite{PhysRevD.64.124012} \COMR{the tBMS transformation}\NEW{tST} at the linear order is shown to be extendable to the horizon of a Kruskal black hole and forms a non-abelian group, it may not be the case for a generic asymptotically flat spacetime.

\paragraph{The choice of the emitting surface}
As discussed in section \ref{sec:HawkingTemp}, the choice of the emitting surface is of paramount importance. We adopt the apparent horizon as it is locally the region with the greatest acceleration still capable of emitting null rays. However, that choice is merely an approximation as the null rays have to reach $\mathcal{I}^+$ globally, i.e. they must originated from the event horizon. Unfortunately to locate the event horizon one must conduct the ray tracing which can only be solved perturbatively or numerically by inserting template forms of $f$. Thus it remains an open question whether the event horizon and the apparent horizon are close enough that we may substitute one by the other at $O(f)$.

\paragraph{Choice of the foliation and its relation to the dressing}
In section \ref{sec:FST_perturb}, we introduce an additional factor $\eta$ for the incoming null vector field $n$ to ensure the close condition $dn=0$ introduced in section \ref{sec:HawkingTemp} where the closeness is required for the global synchronization of the clock. However for local observers such a condition is superficial, one may consider whatever apparatus setup that best suits, e.g. observers synchronized according to the asymptotic Killing vector $n_r$ (the condition we choose in section \ref{sec:FST_perturb} and \ref{sec:exact}).

Notice that a rescaling of $n$ by $\eta$ modifies $\kappa$ by $\kappa \rightarrow \kappa / \eta + \partial_v \eta^{-1} $. In the case of section \ref{sec:main} where $\eta = 1 + f'$, the modification reduces to $\kappa \rightarrow \kappa - \kappa f' - f''$, and by comparing with eq.~\eqref{eq:dressing_tST} we claim that it can be interpreted as an active \COMR{BMS transformation}\NEW{tST} on $\mathcal{I}^+$. While this may appear as an abuse of notion given that in section \ref{sec:main} $f'$ refers to $f'(v)$ whereas in eq.~\eqref{eq:dressing_tST} it corresponds to $f'(u)$, the existence of $\dot p \propto e^{- \kappa u}$ in section \ref{sec:dressing_tST} ($p$ is the ray-tracing function introduced in section \ref{sec:HawkingTemp}) suggests otherwise. %Indeed after relating the rescaled observer velocity $\eta n^r$ to the corresponding $du$ by the Kruskal coordinate $du \propto d \left( \frac{r}{2M} e^{r/2M} \right) $ we have 
Indeed by redefining $\eta = \frac{d\tilde u}{du} \equiv 1 + \tilde f'$ where $\tilde u$ is the observer clock and $u$ is the global clock, and assuming the adiabaticity of $\tilde f'$ in $v$ space near $u = \tau$, we have $\tilde f'_\tau \approx \tilde f^{(1)}_\tau + \tilde f^{(2)}_\tau \left( p \left(u\right) - p \left(\tau\right) \right)$.  %utilize the inverse Legendre transformation to transform the adiabatic parameters $\tilde f'^*$ and $\tilde f''^* \equiv \frac{d\tilde f'^*}{dv}$ to $\tilde f' \equiv \underset{\tilde f''^*}{\sup} \left( \tilde f'^* + \tilde f''^* \frac{dp}{du} u \right)$
Given the adiabaticity of $\kappa$, $p \left( u \right)$ can be approximated as $p \left( u \right) - p \left( \tau \right) \approx \dot p \left( \tau \right) \kappa^{-1} \left( 1 - e^{-\kappa \left( u-\tau \right)} \right)$, and thus $\tilde f_\tau \left( u \right)$ becomes $ \tilde f^{(0)} + \tilde f^{(1)}_\tau u + \tilde f^{(2)}_\tau \dot p \left( \tau \right) \kappa^{-2} \left( \kappa \left( u - \tau \right) - 1 + e^{-\kappa \left( u-\tau \right)} \right)$. While $\tilde f_\tau \left( u \right)$ is not exactly the same as $f_\tau$ in eq.~\eqref{eq:fform}, the difference is minute enough (logarithmic) for us to directly identify one as the other, thus providing a concrete ground for the form chosen in section \ref{sec:dressing_tST}.

\paragraph{Deformation of the apparent horizon}
Although as shown in section \ref{sec:exact} the surface gravity on the apparent horizon is modified by the varying anisotropic incoming null flow, this evidence alone still cannot refute the argument that the modification is induced by the dressing due to a special \COMR{tBMS transformation}\NEW{tST}. To arrive at a definite conclusion we consider the deviation of the Hawking radiation intensity from that emitted by a perfect sphere.

Since the system is perfectly foliated by $v$, the additional lensing experienced by the outgoing null geodesic while traveling toward $\mathcal{I}^+$ with weak lensing approximation can be integrated as $\frac{1}{4M} \left( - \left( \scD^2 + 1 \right) f \left( v \right) + \int_v^\infty dv_1 \scD^2 f' \left( v_1 \right) / \omega\left( \frac{v_1-v}{4M} \right) \right) + O\left( f^2, M' \right)$, where $v$ is the location of the emission and $\omega$ is the Wright omega function, i.e. the inverse of $F(x) = x +\log x$.
The two terms correspond to respectively the apparent lensing induced by the choice of gauge, and a tail of $f'$ suppressed by $1 / r^*$ along the light cone where $r^*$ is the tortoise radius.

Obviously it is different from that of eq.~\eqref{eq:psi}, and thus only when the incoming flow becomes isotropic can both the temperature and the intensity of the Hawking radiation appear isotropic simultaneously. \NEW{This is exactly the same as the CMB weak lensing where the lensing anisotropy induced by the foreground can be isolated from the temperature anisotropy, even if they both originates from the primordial perturbation.} However given that the exact location of the event horizon remains obscure, we can not rule out the possibility that the perturbation of the event horizon radius happens to be the same form (proportional to the one above) as that due to the dressing. Luckily such a scenario is not very persuasive (and will be ignored in the following discussion) as the integral form violates the causality by requiring the entire history of the incoming flow to construct.

\paragraph{Dressing in a time-dependent system} 
%Doppler effect}
%In \cite{Bousso:2017dny}, it has been shown that the dressing factorization which happens to be a coordinate transformation, could remove the soft hair proposed by Strominger et al. However, in our consideration of dressing with a \COMR{time-dependent}\NEW{flow-induced} supertranslation, the process is not guaranteed to be a coordinate transformation anymore. From the derivation in the previous sections, the correlation of Hawking spectrum is modified by the \COMR{time-dependent}\NEW{flow-induced} supertranslation. This result might imply there exist a self-entanglement between Hawking radiation particles, which could be observed experimentally. It gives us a clear guideline to examine whether the time-dependent dressing is equivalent to \COMR{tBMS}\NEW{tST} or \COMR{tST transformation}\NEW{FST}. We have shown that the \COMR{tBMS}\NEW{tST} transformation transforms the metric covariantly with a time-dependent generator of BMS transformation. The physical quantities are not modified accordingly. \COMR{The tST transformation}\NEW{FST}, however, changes the metric such that the surface gravity near the horizon is modified. Thus, we expect that the time-dependent dressing is better described by \COMR{the tST transformation}\NEW{FST}.

As shown in \cite{Bousso:2017dny,Javadinazhed:2018mle}, the dressing (a.k.a. the soft factorization) as a coordinate transformation could separate the soft gravitons from other fields, rendering the \COMR{BMS transformation}\NEW{supertranslation} induced by the energy momentum tensor and the hard gravitons indistinguishable from the soft gravitons.
This discovery invalidates most attempts to explain the information loss paradox by the soft graviton. To dodge the soft factorization we introduce \COMR{the tST transformation}\NEW{FST as an} induced \NEW{transformation} by an incoming anisotropic continuous null flow and compare it with a mimicking coordinate transformation (\COMR{tBMS}\NEW{tST}) in section \ref{sec:dressing_tST} and \ref{sec:exact}. However, we have yet considered to what degree could ``not so soft'' gravitons mimic \COMR{tST transformation}\NEW{FST}.

Given the incapability of the gravitational wave to generate the convergence it is obvious that the Hawking radiation intensity introduced in the previous subsection should serve as the testimony of the incoming flow. With three observables: the proper acceleration (for fixing the gauge), the temperature and the intensity of the Hawking radiation originated from the black hole, an observer on $\mathcal{I}^+$ (without loss of generality with negligible outgoing matter flows and only a central black hole) in principle can distinguish incoming null flows trapped inside the black hole from free-streaming gravitational waves. To testify our argument, however, requires further investigation.

\paragraph{Evasion of the no-hair theorem and the implication to the information loss paradox}

The no-hair theorem is the foundation of the black hole thermodynamics and the precursor to the information loss paradox. To escape the paradox , i.e. to allow information about the Hawking radiation to exist outside of the black hole for the observer to receive, one must forego the no-hair theorem that forbids any measurable independent of the mass, the angular momentum and the charges. The main contribution of this work is to dodge the theorem by introducing varying anisotropic flow that can be measured by the macroscopic properties of the Hawking radiation, or equivalently by hanging a rope near the horizon and measure the perceived force.% However, given how the information is encoded in eq.~\eqref{eq:psi} one would need access to 

We have to emphasize that only the classical properties of the spacetime (e.g. the Hawking radiation intensity, the lensing potential and the surface gravity) are considered in our work. To retrieve the information entangled with the Hawking radiation, however, is a completely different feat and requires more throughout analysis of the horizon dynamics. One particular way of approaching this issue is to generalize the black hole thermodynamics \cite{PhysRevLett.92.011102} by incorporating the spatial-temporal behavior of the Hawking radiation, and reverse-engineer the interaction between the impulsing Hawking radiation and the responding radiation from the response function.

%The black hole soft hair in dynamical spacetime is an essential assumption in the study of the soft hair proposal to the information loss paradox. The soft hair is related to the collapsing matter such that it could encode the information of black hole formation and evaporation process. In this article, we have found that by only assuming a dynamical spacetime is not sufficient to evade the no-hair theorem. That is, the back reaction of spacetime from matter flow would affect the supertranslation and make it a time-dependent function. After considering this modification, we have obtained observable soft hair near the horizon.

%To solve the information loss paradox, a gateway is to find a new physical observable in addition to ADM charges. In this work, we have found a minimal modification on the Hawking temperature during the black hole accretion process. We have imposed a \COMR{time-dependent}\NEW{flow-induced} supertranslation soft hair and have found that the surface gravity on the horizon is modified by the back reaction of anisotropic energy flow. By applying the \COMR{time-dependent}\NEW{flow-induced} supertranslation soft hair, we have also found that the spectrum of Hawking particle created by a black hole would be modified in consequence. These results might give us an opportunity to claim that the \COMR{time-dependent}\NEW{flow-induced} supertranslation soft hair could encode part of the information during the black hole evaporation process.  

%%%%%%%%%%%%%%%%%%%%%%%%%%%%%%%%%%%%%%%%%%%%%
%%%%%%%%%%%%%%%%%%%%%%%%%%%%%%%%%%%%%%%%%%%%%

\section{Conclusions}
\label{sec:conclusion}

We generalize the setup of \cite{Hawking:2016sgy,Chu:2018tzu} where an anisotropic shock wave falls into the central Vaidya black hole and generates BMS charges at the linear order, to a setup with an incoming continuous anisotropic null flow generating BMS charges on the fly. In the process we realize the existence of an asymptotic coordinate transformation other than the BMS transformation, which serves as the dressing of the hard particles in the soft factorization procedure. Together they form a linear response relation with the energy momentum tensor and can be shown to be the exact BMS charge generator valid well beyond the horizon, which is associated with a current flowing directly into the black hole. We also carry out the effect of the dressing on the Hawking radiation, which happens to be equivalent to the Doppler effect and the Unruh effect of a non-stationary observer at the future null infinity.

Furthermore, we find a modification to the surface gravity as shown in eq.~\eqref{exsur} up to the linear order, an effect previously gone unnoticed due to the vanishing of the linearized tidal Love number in a 4$-$d black hole system. This modification depends on the exponentially weighted average of the anisotropic energy flow, an encoding different from that originated from the usual line-of-sight integration involving the tortoise coordinate, and thus is unlikely to be originated from the dressing on the horizon. This new effect can be regarded as an access to the BMS charges on a black hole without the intervene of the no-hair theorem, and could be the first step toward the resolution of the information loss paradox, with lots of possible extensions for further study.

\section*{Acknowledgments}
We appreciate the discussions with Che-Yu Chen, Pei-Ming Ho, Keisuke Izumi, Misao Sasaki, and W. G. Unruh. This work is supported by National Center for Theoretical Sciences (NCTS) of Taiwan, Ministry of Science and Technology (MOST) of Taiwan, and the Leung Center for Cosmology and Particle Astrophysics (LeCosPA) of National Taiwan University.

%%%%%%%%%%%%%%%%%%%%%%%%%%%%%%%%%%%%%%%%%%%%%
%%%%%%%%%%%%%%%%%%%%%%%%%%%%%%%%%%%%%%%%%%%%%

%\appendix
%\begin{appendices}
%
%
%\section{Insensitivity of the horizon to a non-covariant transformation in a single-horizon system}
%\label{sec:appendices}
%In a system with only one horizon $\mathcal{H}$, i.e. only one hypersurface with a vanishing expansion rate $\theta_L$, there exists a linear relation $\theta_L = a \left( r - r_\mathcal{H} \right) + O \left( (r - r_\mathcal{H})^2 \right)$, where $a$ and $r_\mathcal{H}$ are two parameters.
%
%
%\end{appendices}

%\newpage
%~
%\newpage
~
\newpage

\bibliographystyle{JHEP}

\bibliography{bibliography}

\end{document}